\documentclass[11pt]{article}
\usepackage{a4wide}
\usepackage{amsmath}
\usepackage{amssymb}
\usepackage{natbib}
\usepackage{color}
\usepackage{bm}
\usepackage{graphics,graphicx}
\usepackage{url}
\usepackage{hyperref}
\usepackage{pdfpages}
\usepackage{multirow}
\makeatletter

\newcommand{\para}{\bigskip\noindent}

\newcommand{\bbeta}{\bm{\beta}}
\newcommand{\boldeta}{\bm{\eta}}

\newcommand{\blambda}{\bm{\lambda}}

\newcommand{\be}{\begin{equation}}
\newcommand{\ee}{\end{equation}}
\newcommand{\TITLE}[1]{{\center{\bf#1}\vskip30pt}}

\newcommand{\AFFILIATION}[1]{{\small\center {\it #1} \\}}
\newcommand{\AUTHORS}[1] {{\small\center{#1}\\}}

\begin{document}
\TITLE{Think before you shrink: Alternatives to default shrinkage methods can improve prediction accuracy, calibration and coverage}

\AUTHORS{Mark A. van de Wiel$^1$, Gwena\"el G.R. Leday$^2$, Jeroen Hoogland$^1$, Martijn W. Heymans$^1$, Erik W. van Zwet$^3$, Ailko H. Zwinderman$^1$}

\AFFILIATION{$^1$Dept of Epidemiology and Data Science, Amsterdam Public Health research
institute, Amsterdam University Medical Centers, Amsterdam, The Netherlands; Biometris, Wageningen University and Research, Wageningen, The Netherlands;
$^3$Dept of Biomedical Data Sciences, Leiden University Medical Center, Leiden, The Netherlands
}

\begin{abstract}
While shrinkage is essential in high-dimensional settings, its use for low-dimensional regression-based prediction has been debated.
It reduces variance, often leading to improved prediction accuracy. However, it also inevitably introduces bias, which may harm two other measures of predictive performance: calibration and coverage of confidence intervals.
Much of the criticism stems from the usage of standard shrinkage methods, such as lasso and ridge with a single, cross-validated penalty. Our aim is to show that readily available alternatives can strongly improve predictive performance, in terms of accuracy, calibration or coverage. For linear regression, we use small sample splits of a large, fairly typical epidemiological data set to illustrate this. We show that usage of differential ridge penalties for covariate groups may enhance prediction accuracy, while calibration and coverage benefit from additional shrinkage of the penalties. In the logistic setting, we apply an external simulation
to demonstrate that local shrinkage improves calibration with respect to global shrinkage, while providing better prediction accuracy than other solutions, like Firth's correction. The benefits of the alternative shrinkage methods are easily accessible via example implementations using \texttt{mgcv} and \texttt{r-stan}, including the estimation of multiple penalties. A synthetic copy of the large data set is shared for reproducibility.
\end{abstract}

\section{Introduction}

Shrinkage has become a standard technique in statistics to counter over-fitting in regression models. In particular in high-dimensional settings, with the number of variables $p$ larger than sample size $n$, application of shrinkage is necessary to obtain parameter estimates and predictions. In low-dimensional settings, the benefit of shrinkage depends on the $n:p$ ratio, as well as the main purpose of the analysis, e.g. prediction, parameter estimation, variable selection or causal effect estimation. Here, we focus on prediction. In general, the consensus seems to be that, on average, shrinkage improves prediction accuracy on test data from the same population as the training data \cite[]{van2020regression, riley2021penalization}. This is no surprise as the penalty parameter is usually tuned, e.g. by cross-validation, to maximize out-of-bag prediction accuracy. Hence, it will adapt to the $n:p$ ratio: if large, little shrinkage is necessary; if small, more shrinkage is required. The use of shrinkage, however, comes at a price. It inevitably biases the parameter estimates, which in turn may lead to bad calibration of the prediction \cite[]{van2020regression}. Moreover, the penalty parameter can be rather instable \cite[]{riley2021penalization}, and such instability is often not communicated when presenting resulting models. Most empirical results that support these critiques on shrinkage are based on standard penalization techniques like lasso and ridge. While these issues are to some extent intrinsic to any shrinkage method, we show that they can be substantially alleviated by using alternatives that are only slightly more advanced. Therefore, our aim is to show a broad audience of applied statisticians and epidemiologists that some `non-standard' shrinkage methods may be very useful for typical epidemiological studies that target multivariable prognostic models.

\para
In terms of shrinkage we focus mainly on several variants of ridge regression. We do briefly discuss comparison of prediction accuracy with classical lasso and stepwise selection (which could be regarded as an extremely bimodal type of shrinkage) for the main data example, showing the latter two are inferior to the ridge variations for this purpose. We emphasize that the latter two cater mainly for variable selection, rendering the comparison with ridge-type models somewhat unfair. Therefore, we focus on the latter.

\para
Our setting is regression, where shrinkage is either effectuated by using a penalty for the parameters or by using an informative prior on these parameters in a Bayesian setting.
In the latter case, the scale parameter of the prior distribution acts effectively as a penalty parameter, as the mean of the prior is usually set to zero.
First, we explore differential shrinkage by the use of different penalties for (groups of) variables. We argue that in many settings, it is in fact unreasonable to assume the same penalty for all variables. We show that both prediction accuracy and calibration may be greatly enhanced by using differential penalties. In addition, the use of differential penalties provides a more objective solution to down-weighting a set of variables than simply discarding it prior to the analysis, as is sometimes suggested as a (partial) solution for a low $n:p$ ratio \cite[]{van2020regression}. Importantly, these penalties can be estimated automatically without (time-consuming) cross-validation in both the classical and full Bayesian setting, using \texttt{R}-packages \texttt{mgcv} \cite[]{wood2011fast} and \texttt{R-Stan} \cite[]{rstan}, respectively.

\para
Global shrinkage methods, such as ridge, penalize all parameters equally. If the primary aim is to optimize prediction accuracy, we show that these methods perform well, in particular when one allows for differential shrinkage. Calibration, however, is often bad, as the reduced variability is countered by increased bias. While this is partly inevitable, we show that local shrinkage \cite[]{gelman2006prior, polson2012half}, which penalizes (groups of) variables differently,  may improve calibration considerably. This comes at the cost of some loss in prediction accuracy as compared to global shrinkage, but local shrinkage still outperforms ordinary least squares (OLS) by a fair margin, while being competitive in terms of calibration.

\para
The reported instability of the penalty parameter(s) \cite[]{riley2021penalization} may affect the uncertainty of the predictions.
Unlike in settings with cross-validated or fixed penalty parameter(s), uncertainty propagation of the penalty parameter(s) is accomplished very naturally in a Bayesian hierarchical shrinkage model. Hereby, it may aid in communicating appropriate prediction uncertainties by improving (frequentist) coverage of their credibility intervals. Moreover, the additional regularization of the penalty parameter(s) may increase stability of their estimates, in particular in the presence of multiple penalties. In the frequentist paradigm penalty parameter uncertainty is not explicitly modeled. \cite[]{marra2012coverage} show, however, that modifying the classical intervals such that the covariance matrix of the regression coefficients matches with the Bayesian one (which includes the additional prior uncertainty of those coefficients) improves coverage. Here, we also consider those `unconditional' intervals, as implemented in \texttt{mgcv} .

\para
Our main strategy to compare various methods is to divide a large ($N = 21,570$), fairly typical epidemiological study into many subsets of sizes $n=50$ and $n=200$.
We study systolic blood pressure (SBP) as outcome, and use twelve variables to predict SBP. We add five random noise covariates to make sure that some covariates are not linked to the outcome at all, resulting in $p=17$. The OLS results on the entire set are used as benchmark, as these estimates are unbiased and very precise due to the large sample size. Then, the subsets are analysed with various shrinkage methods and compared to the benchmark to evaluate prediction accuracy (using out-of-bag test samples), calibration and
coverage of the prediction interval.
We also analyse two simulated data sets: an external one for the logistic setting \cite[]{van2020regression}, which focuses on calibration; and a second one serving as the following introductory, motivating example.

\subsection{Introductory example}
Our motivating simulation example is loosely inspired by Model A in \cite{riley2021penalization}, which relates seven covariates to SBP. Treatment (yes/no) is one of the covariates. The results of OLS in \cite{riley2021penalization} show a strong treatment effect as compared to the other six parameters. Such a scenario may be fairly common in clinical studies. Therefore, we simulate regression coefficients that adhere to this scenario:
$\beta_1 = \beta_2 = \beta_3 = -0.05, \beta_4 = \beta_5 = \beta_6 = 0.05$, $\beta_{\text{Treat}} = -0.25$ and error variance $\sigma^2=1$.
We simulate 1,000 data sets of size $n=50$ from the linear model. To each of these, we apply standard ridge shrinkage, as in \cite{riley2021penalization}, and two-penalty ridge shrinkage (ridge\_2) allowing differential shrinkage for the treatment parameter. We argue the latter is reasonable, as treatment is an intervention and therefore a different type of covariate than the others. In the Bayesian setting (Bay\_2) the square-roots of the two inverse penalties (i.e. standard deviations) are endowed with a standard $\text{C}^+(0,1)$ (half-Cauchy) prior \cite[]{gelman2006prior}. Table \ref{sim} shows the evaluation results in terms of mean squared error of the prediction (MSEp) and mean coverage of the 95\% confidence intervals of the predictions, as evaluated on a large test set ($n_{\text{test}} = 1,000$).

\begin{table}[h]
\begin{center}
\begin{tabular}{|l||l|l||l||l|}\hline
Method &  MSEp, mean  &  MSEp, 90\% & Cover\\\hline
ridge & 0.096 & 0.157 & 0.844 \\
ridge\_2 & 0.079 & 0.150 & 0.898\\
Bay\_2 & 0.068 & 0.131 & 0.969
\\\hline
\end{tabular}\caption{Results from simulation example. MSEp: mean squared error of the prediction; mean and 90\% quantile across data sets; Cover: mean coverage of the 95\% confidence intervals of the predictions (target: 0.95)}\label{sim}
\end{center}
\end{table}
\noindent
The results clearly show the benefit of two-penalty shrinkage as compared to standard ridge:  ridge\_2 and Bay\_2 decrease the mean MSEp by 17.7\% and 29.2\%, respectively. Most striking is the difference in interval coverage.
While Bay\_2 is somewhat conservative it is much closer to the target than ridge and ridge\_2. We repeated the simulation for a setting that is more beneficial to global ridge shrinkage: all regression parameters equal $\pm 0.1$. Then, prediction accuracies of all three methods are very similar, implying the cost of using one redundant penalty parameter, as for ridge\_2 and Bay\_2, is limited. The interval coverage of Bay\_2 is still superior to that of ridge (ridge\_2) as it averages to 95.7\% versus 86.0\% (87.8\%).

\para
Below we study various frequentist and Bayesian shrinkage methods in more detail for real data in the linear regression setting. We evaluate methods by considering prediction accuracy, coverage of the confidence intervals of the predictions, and calibration of the predictions. As the latter is particularly also a concern in binary prediction problems \cite[]{van2020regression} we address this in an external simulation setting. To enhance reproducibility of our results we provide i) a synthetic copy of our primary data set, for which we show it renders qualitatively the same results as the real data; and ii) all code to run and evaluate the models, including the simulations. The latter should also assist researchers to use some of the `non-standard' solutions we propose for their own data. We end with a discussion, which includes recommendations and several extensions.

\subsection{Data}

The main data we use throughout the manuscript is obtained from the Helius study \cite[]{helius2017}. In a linear regression context, we study response $Y$: systolic blood pressure (SBP), as a function of covariates $X$: age, gender, BMI, ethnicity (5 levels), smoking (binary), packyears, coffee (binary), glucose (log), cholesterol, rendering twelve covariates after dummy coding the nominal covariate. We apply minimal preprocessing to the data: only 2.7\% of the samples had at least one missing value; these samples were removed, rendering a sample size $N= 21,570$. Continuous covariates were standardized, as this is common practice before applying shrinkage for the penalty to have the same effect on all corresponding regressing parameters. For binary covariates, the standardization itself may become unstable for small data sets, in particular when the classes are unbalanced. This may hamper generalization to test settings. Therefore, we opted for a default -1, 1 coding, as this standardizes a balanced binary covariate. For our data, we compared results with those from complete standardization. Differences were small, but marginally better for the proposed coding. Finally, we added five independent standard normal noise covariates. Hereby, we are sure to include some covariates completely unrelated to the outcome, SBP. Hence, the total number of covariates equals 17.

\para
We chose this data set for various reasons. First, it addresses a fairly standard, and well-known prediction problem with an interesting mix of covariates (binary, nominal and continuous). Second, its large sample size allows to a) use the OLS estimates from the entire set as the benchmark, because these have very small standard errors (typically $\approx 0.01$); and b) to split the set in many independent subsets with sample sizes as often encountered in clinical studies, such as $n=50, 200$. This enables us to  evaluate various (shrinkage) methods on many real, relatively small sample data sets. Third, from applying OLS to the entire data set, we have $R^2 = 0.34$, which is neither trivially low nor high. The data cannot be shared, but we provide a synthetic copy which qualitatively renders the same results as those presented here.

\section{Methods}
All methods fit the linear model $Y_{i} = \beta_0 + X_i\bbeta = \beta_0 + \sum_{j=1}^{17}\beta_j X_{ij} + \epsilon_{i}, \epsilon_{i} \sim N(0,\sigma^2)$, with $i=1, \ldots, n$,
$\bbeta = (\beta_1, \ldots, \beta_{17})$ and covariates $X_i = (X_{i1}, \ldots, X_{i17})$.
Before discussing the methods, we state the evaluation criteria, which all focus on prediction.
\subsection{Evaluation criteria} First, note that the `true' value of $\bbeta$ is obtained from applying OLS to the entire data set, as this rendered estimates with extremely small standard errors. Training sets $T^s$ and their complementary test sets $T'^s$ are indexed by $s$.  The true value of the prediction for any (test) sample $i$ with covariates $X_i$ then equals $\eta_i = \beta_0 + X_i\bbeta$. We evaluate the methods on three criteria:
\begin{enumerate}
  \item \emph{Prediction accuracy}. We use the mean squared error of the predictions (MSEp). For a given subset $s$ this is defined as
   \begin{equation}\label{msepred}
   \text{MSEp}^s = 1/|T'^s| \sum_{i \in T'^s} (\eta_i - \hat{\eta}^s_i)^2,
   \end{equation}  with, for test sample $i$, $\hat{\eta}^s_i = \hat{\beta}_0^s + X_i\hat{\bbeta}^s$, where $\hat{\beta}_0^s$ and $\bbeta^s$ are estimated from samples in training set $T_s$.
   As we have the `true' predicted values $\eta_i$ in this setting, we use those in \eqref{msepred} instead of $y_i$, which is often used when $\eta_i$ is not avalaible, rendering the prediction squared error. Of note: we checked that the latter leads to the same conclusions as the former.
  \item \emph{Calibration of the predictions}. As we are in the setting of knowing the true values, we use the term `calibration' as it is used in measurement technology: we quantify how close the predictions (which are derived from measurements) are to an accurate benchmark, in this case the true values. For this, we regress predictions $\hat{\eta}^s_i$ on their true values $\eta_i$ per subset $s$, including an intercept as well. The slope of this regression, which we term `cslope', serves as a calibration metric as it should be close to 1, and quantifies the possibly biasing effect of shrinkage across the range of predictions.
      Note that this deviates from what has become known as the `calibration slope' in the literature, which results from regressing the observed values on the predictions. The latter provides an assessment of the overall agreement of observed values and predictions, but has been claimed to be a misnomer, partly because it is (also) a measure of spread of the predictions \cite[]{stevens2020validation}.

  \item \emph{Uncertainty of the predictions}. We evaluate this by the mean coverage of the confidence intervals of the predictions $\boldeta = (\eta_1, \ldots, \eta_{N})$. In the classical penalized ridge regression setting we use the 95\%  Gaussian Wald intervals with standard errors corrected for prior uncertainty of $\bbeta$, as derived in \cite{marra2012coverage}. For Bayesian methods, we simply use the (2.5\%, 97.5\%) quantiles of the posteriors to create 95\% credible intervals for the predictions $\eta_i$. Note that confidence intervals of predictions $\boldeta$ do not include the measurement error $\epsilon_{ij}$ as prediction intervals do. For simplicity, we decided to focus on the former as these more directly relate to the shrinkage of regression coefficients $\bbeta$, and less so to how the error variance $\sigma^2$ is estimated.
\end{enumerate}

\subsection{Standard solutions}
We study OLS, stepwise selection, lasso and ridge as standard solutions with no or just one global penalty parameter. The intercept is not penalized. Note that step is also a shrinkage method, but of an extreme nature: it shrinks completely to 0 or not at all, hence somewhat similar to the use of a spike-and-slab prior in a Bayesian model.
Here, step is the standard \texttt{step} implementation in \texttt{R}, using forward-backward selection and AIC to select a model.
We use lasso as implemented in \texttt{glmnet}, using \texttt{lambda.min} resulting from 10-fold cross-validation (CV). For ridge, we estimated penalties using marginal likelihood optimization as implemented in \texttt{mgcv} \cite[]{wood2011fast}, which, as opposed to \texttt{glmnet}, provides confidence intervals \cite[]{marra2012coverage}. Results from 10-fold CV were very similar, and hence not shown.

\subsection{Multi-penalty solutions}
Multi-penalty solutions allow the use of different penalties for groups of covariates.
We focus on ridge-type solutions for three reasons: a) our focus lies on prediction and not on variable selection; b) standard software implementations such as \texttt{mgcv} allow efficient estimation of the penalties; and c)  standard ridge performs relatively well (see Supp Fig \ref{standard}).
In some settings, the use of covariate groups is very natural, e.g. when some covariates represent similar entities (e.g. genes), or when main effects plus interactions are included.
In other cases, like ours, choice of the covariate groups implies a level of subjectivity. Therefore, we also assess the performance of multi-penalty approaches when the covariate groups are chosen randomly; that is, when the prior information used to form the covariate groups is useless. Besides differential penalization another option is not to shrink one (group of) covariates, for example because there is substantial evidence that these effect the outcome. We will explore this option as well.

\para
We now define the covariate groups that were used:
\begin{itemize}
  \item Two groups, $G=2$. The first 3 covariates (age, gender, BMI) are one group, supposedly because these are known to relate to SBP from previous studies. The other 14 covariates are in the second group. Two options are explored: ridge\_2 penalizes both groups, separately, whereas ridge\_2un leaves the first group unpenalized.
  \item Three groups, $G=3$, ridge\_3. As above, but the nominal covariate, ethnicity, is a seperate group as this is a nominal covariate with five levels; here, `Dutch' is chosen as the baseline level.
  \item Two random groups. Three covariates are randomly picked to belong to group 1, the other 14 belong to group 2. Randomization is repeated for each training data set. Referred to as ridge\_2r and ridge\_2unr, the latter corresponding to leaving the smallest group of covariates unpenalized.
\end{itemize}

\para
\cite{riley2021penalization} and others have reported the instability of the penalty parameters for standard solutions like ridge. This instability is likely
to increase when estimating multiple penalty parameters, as less data per penalty parameter is available then. Bayesian solutions may be a worthwhile alternative to shrink these penalties with an additional prior to counter such instability. Moreover, the extreme case: one penalty per parameter, referred to as local shrinkage, may benefit calibration, as illustrated further on.

\subsection{Bayesian solutions}
Shrinkage is a very natural concept in Bayesian statistics. Depending on the type of shrinkage, frequentist and Bayesian solutions may be very similar.
For example, the frequentist ridge and lasso estimates of $\bbeta$ are equal to the posterior mode estimate of $\bbeta$ when using
a Gaussian or Laplacian prior with a fixed precision parameter that relates proportionally to the penalty parameter for fixed sample size.

\para
Bayesian methodology, however, can model the variability of the penalty parameter, which holds several promises. First, it counters potential instability of estimates of penalty parameter(s) by using a (weakly) informative prior for it. Second, it propagates uncertainty of penalty parameter(s). This follows from the fact that for a single $\beta$ and random penalty parameter $\lambda$ we have $V(\beta) = E_{\lambda}[V(\beta|\lambda)] + V_{\lambda}[E(\beta|\lambda)].$ Finally, it allows local shrinkage, either by group, or even per covariate. Note that, just as for OLS, an advantage of local shrinkage is that the results are less sensitive to (differences in) scale and effect sizes, as the local penalty can adapt.

\para
Extensions to Bayesian ridge regression differ in how they allow the penalty parameters to vary, a priori.
Table \ref{priors} specifies the priors that we used.
\begin{table}[h]
\begin{center}
\begin{tabular}{|l|l|l|l|}\hline
Name & Prior $\beta$ & Prior penalty $\lambda$ & Type of shrinkage\\\hline
Bay\_EB & $\beta_k \sim N(0,\sigma^2\lambda^{-1})$ & $\lambda = \hat{\lambda}_{\text{EB}}$ & global, fixed\\
Bay\_IG & $\beta_k \sim N(0,\sigma^2\lambda^{-1})$ & $\lambda^{-1} \sim \text{IG}(0.001,0.001)$ & global, vague\\
Bay\_glo & $\beta_k \sim N(0,\sigma^2\lambda^{-1})$ & $\lambda^{-1/2} \sim \text{C}^{+}(0,1)$ & global, weakly informative\\
Bay\_2 (3) & $\beta_k \sim N(0,\sigma^2\lambda^{-1}_{g(k))}$ & $\lambda^{-1/2}_g \sim \text{C}^{+}(0,1)$ & grouped, weakly informative\\
Bay\_loc & $\beta_k \sim N(0,\sigma^2\lambda^{-1}_k)$ & $\lambda^{-1/2}_k \sim \text{C}^{+}(0,1)$ & local, weakly informative\\\hline
\end{tabular}\caption{Priors}
\end{center}
\end{table}\label{priors}

\noindent
Here, $g(k)$ denotes the map of covariates to groups: $\{1, \ldots, 17\} \rightarrow \{1, \ldots, G\}$, with $G=2,3$. EB refers to `empirical Bayes', meaning that
the penalty parameter is considered fixed and estimated by maximizing marginal likelihood. IG is the conjugate inverse-gamma prior, and $\text{C}^{+}(0,1)$ is the half-Cauchy prior (see Supp Fig \ref{halfcauchy}) for the standard deviation(s), which has fairly heavy tails and was advocated as a good default prior by \cite{gelman2006prior, polson2012half}. Moreover, $\sigma^2$, the error variance in the linear model (set to 1 in above's prior for the logistic), is endowed with either a Jeffrey's prior or a vague inverse gamma (results very similar). It is standard practice to include $\sigma^2$ in the prior of $\beta_k$ in Bayesian linear regression.
Since we observed that the global versions of Bay\_EB and Bay\_IG performed very similarly to their frequentist counterpart (ridge; see Supp Fig \ref{ridgevariations}), we primarily focus on the C$^+$ prior for the grouped and local settings.

\section{Results}
\subsection{Prediction accuracy}
We first consider prediction accuracy, as measured by the mean squared error of the predictions, MSEp \eqref{msepred}.
In all figures, results are based on 400 training subsets of sizes $n=50, 200$, which due to the size of the entire data set show no or little overlap in samples. Predictions are evaluated on complementary test sets. Below, we discuss the results for several comparisons.

\para
\textbf{Standard solutions}: We first compare the standard methods. Suppl. Fig. \ref{standard} depicts the performances. For $n=50$, we observe that ridge performs better than lasso, which is substantially better than stepwise selection and OLS. For $n=200$ the gap between ridge and lasso becomes much smaller, and so do the other gaps.

\begin{figure}[h]
\centering
\includegraphics[scale=0.9]{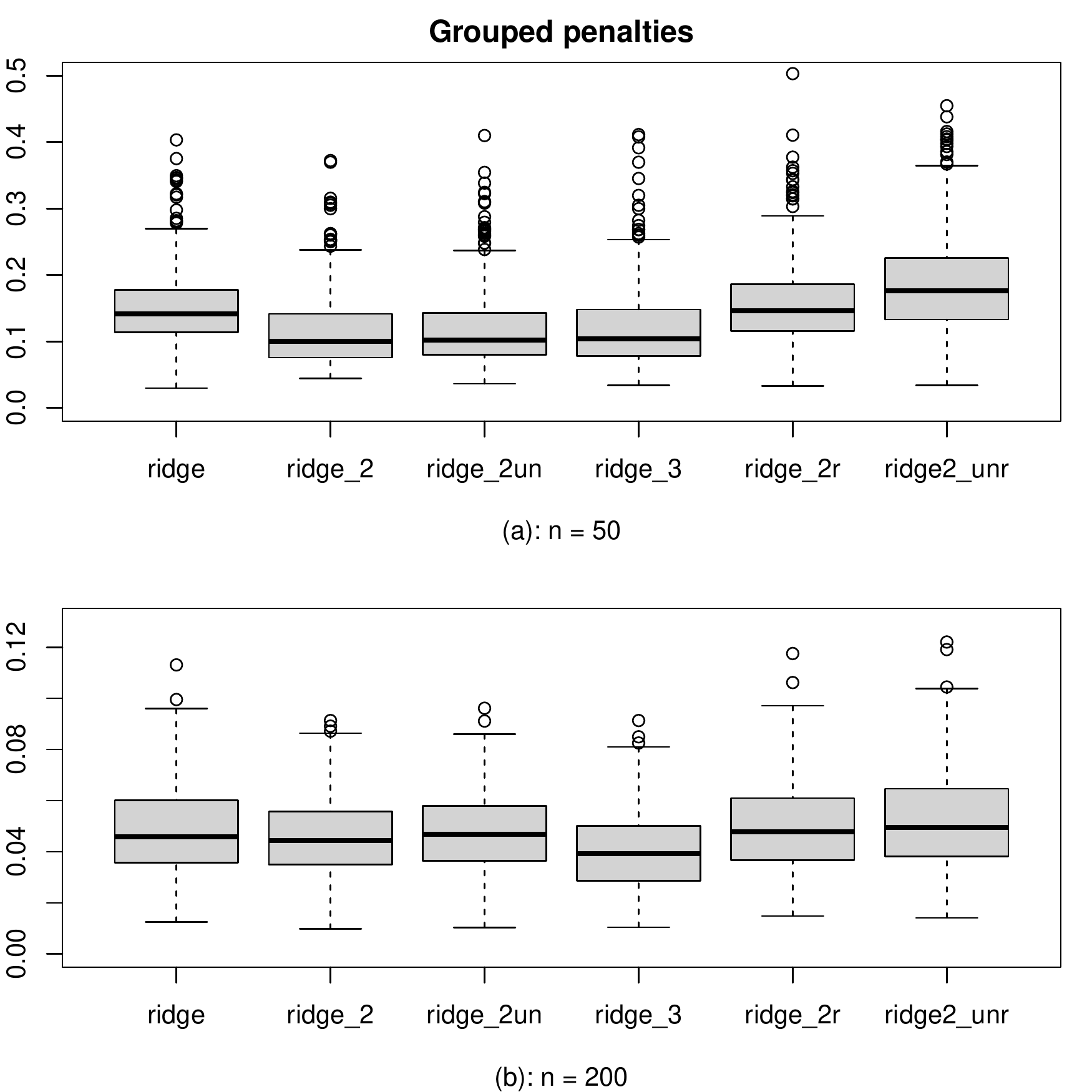}
\caption{MSEp (y-axis) for grouped penalization across 400 subsets. Digital suffix denotes number of covariate groups, suffix `un' denotes one unpenalized group,
suffix `r' denotes random groups}\label{groups}
\end{figure}

\begin{figure}[h]
\centering
\includegraphics[scale=0.5]{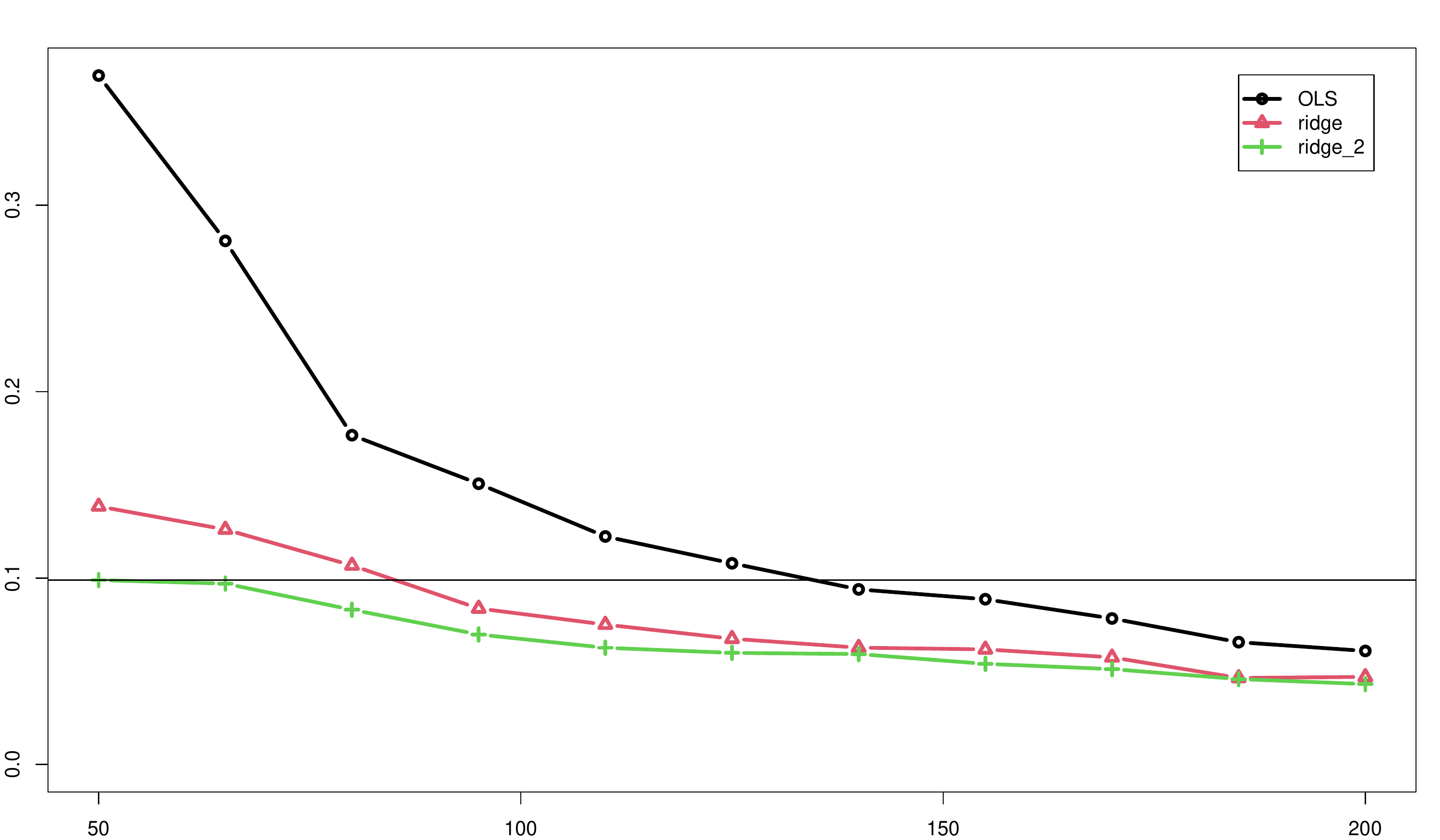}
\caption{Median MSEp (y-axis) across 400 subsets (y-axis) for various sample sizes (x-axis) for OLS, ridge and 2-group ridge}\label{samplesize}
\end{figure}

\para
\textbf{Multi-penalty solutions}:
Fig \ref{groups} shows that use of multiple penalties can improve prediction accuracy (lower MSEp). For n=50, we observe a substantial improvement for $G=2$ penalties, likely caused by the large difference in penalties for the two groups (Supp Fig \ref{varpen2}; first and third boxplot).
For n=200, the improvement for $G=2$ is marginal, but the larger sample size benefits the estimation of one more penalty rendering improved accuracy for $G=3$.
  In this setting ridge\_2un, which does not penalize the first three covariates, performs on par with ridge\_2. This is reasonable, because these covariates are relatively strong, so need little penalization. The figure also shows, however, that when the two groups are assigned at random ridge\_2unr performs inferior to ridge\_2r (with suffix `r' denoting the random group setting) as it cannot adapt when the unpenalized group is relatively weak, and would hence have benefitted from penalization. Importantly, also observe that even when the groups are random, ridge\_2r performs - on average - (nearly) on par with ridge, again due to the adaptive penalties. The improvement with the multi-penalty approach becomes very tangible in Fig. \ref{samplesize}. It shows that OLS and ridge need many more samples to achieve the same median MSEp as ridge\_2, f.e. approximately 135 and 85, respectively, versus 50.

\para
\textbf{Adding Bayesian solutions}:
We first compare the standard (empirical) Bayesian ridge regression solutions, Bay\_EB and Bay\_IG,  with their frequentist counterpart, ridge. Suppl. Fig. \ref{ridgevariations} shows very little difference in MSEp between those methods. This is expected as for both methods the posterior mode estimate coincides with the frequentist estimate. Small differences are due to the use of posterior mean instead of posterior mode as a summary for the predictions.

\begin{figure}[h]
\centering
\includegraphics[scale=0.9]{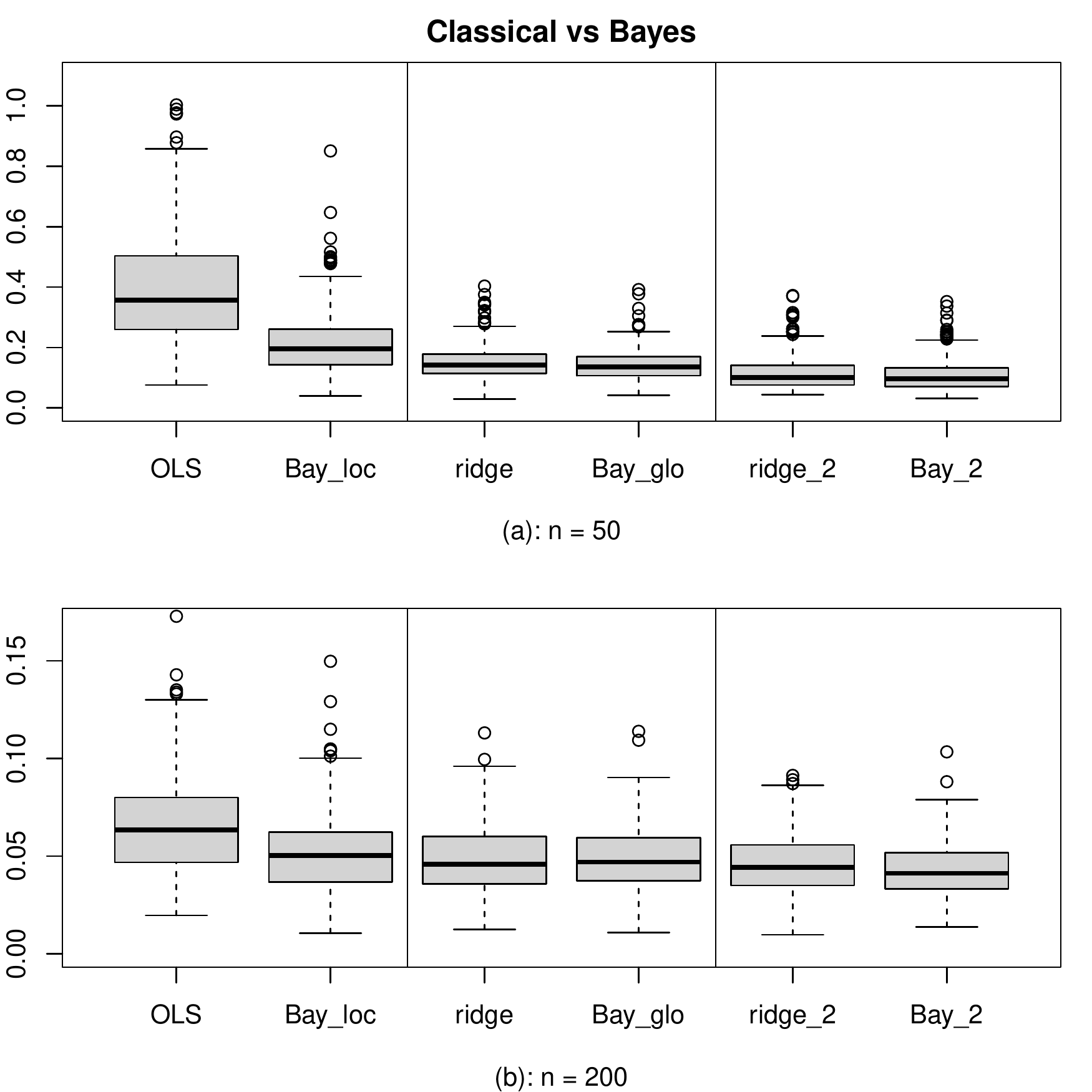}
\caption{MSEp (y-axis) for classical methods and their Bayesian counterpart across 400 subsets. Digital suffix denotes number of covariate groups}\label{freqBay}
\end{figure}

\para
Fig. \ref{freqBay} compares each of the three Bayesian methods with C$^+$ priors to their frequentist counterparts. Here, Bay\_loc is contrasted with OLS as both methods do not imply any grouping on the covariates. Likewise, Bay\_2 is contrasted with ridge\_2 (2 groups) and  Bay\_glo is contrasted with ridge (1 group). For the latter two comparisons we observe that differences in MSEp are minor. A more striking difference is observed for OLS vs Bay\_loc. While the latter cannot compete with the global shrinkage methods (for $n=50$), it does perform much better than OLS. Apparently, the default shrinkage of the normal-C$^+$ prior substantially stabilizes the $\beta$ estimates in such small sample size settings. The effect of the local normal-C$^+$ prior for $\beta$ is depicted in Supp. Fig. \ref{priorbeta} for two $n=50$ subsets with extreme OLS estimates of one of the most important coefficients, $\beta_{\text{BMI}}$. We observe that the normal-C$^+$ prior of Bay\_loc shrinks the extreme OLS estimates in the right direction.

\subsection{Calibration}
\begin{figure}[h]
\centering
\includegraphics[scale=0.43]{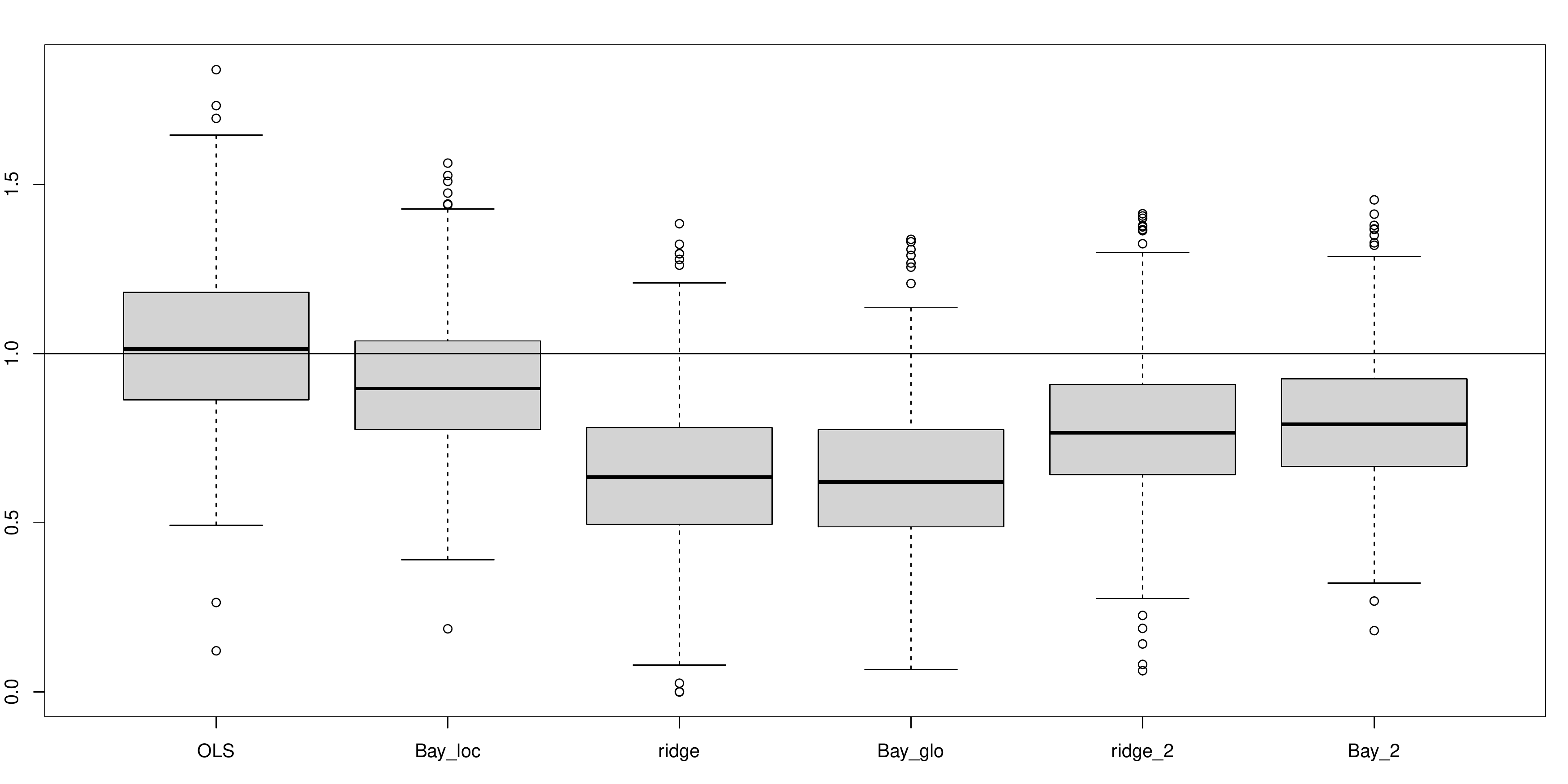}
\caption{Calibration: cslopes (y-axis) across all 400 subsets of size $n=50$}\label{calibration}
\end{figure}

Fig. \ref{calibration} shows the `cslopes' resulting from regressing test sample predictions $\hat{\eta}^s_i$ against true $\eta_i$ for all subsets $s$ for OLS, Bay\_loc, ridge, Bay\_glo, ridge\_2 and Bay\_2. We observe that OLS and Bay\_loc outperform the other methods on calibration. OLS and Bay\_loc are competitive: the first being unbiased, but with larger variability of the slopes; in fact, the root MSE of the slope estimates (as compared to 1) are comparable: 0.249 (OLS) vs 0.232 (Bay\_loc). The bias of ridge and Bay\_glo is evident; shrinkage compresses the slopes. Differential penalization using two covariate groups improves results as compared to global regularization, with a small edge for Bay\_glo2 w.r.t. ridge\_2 in terms of decreased variability likely due to the extra regularization of the penalties. Differences in calibration performance are somewhat smaller for $n=200$ (Supp. Fig. \ref{calibrationn200}).

\para
To elaborate on the inferior calibration of ridge and  Bay\_glo compared to OLS and Bay\_loc consider the estimated coefficients of a strong covariate (in the entire study), BMI, from the $n=50$ subsets. Supp. Fig. \ref{bmicoefs} clearly shows that ridge and  Bay\_glo over-shrink $\beta_{\text{BMI}}$ due to the common shrinkage factor shared with the weaker covariates, whereas Bay\_loc strikes a good balance between a small amount of bias and reducing variability compared to the OLS estimates.

\subsection{Quantifying uncertainty: confidence interval of predictions}
\begin{figure}[h]
\centering
\includegraphics[scale=0.43]{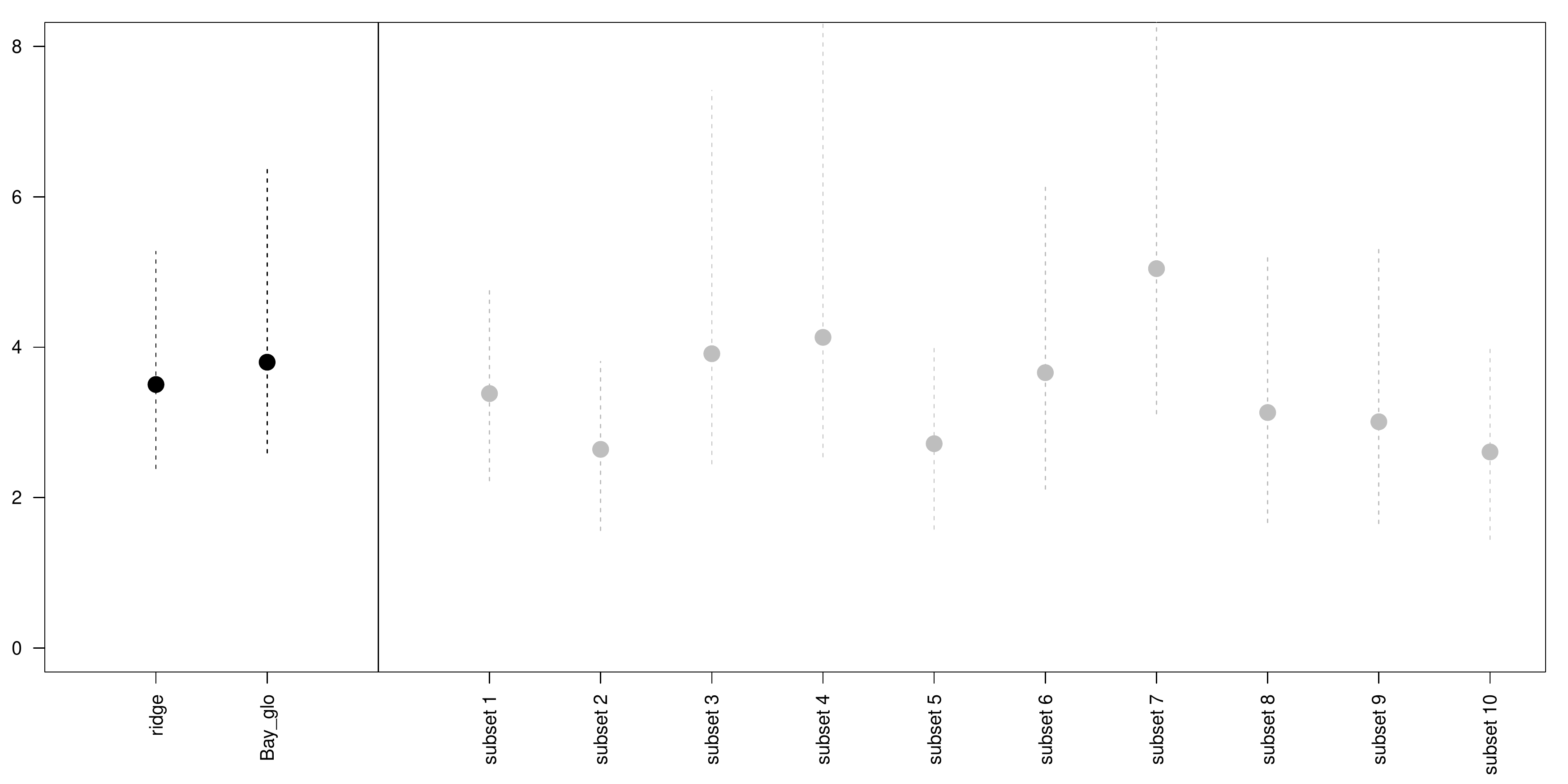}
\caption{Variability of ridge penalties. Y-axis: ridge penalties on natural log-scale. Left panel (black):  ridge penalties estimated by
\texttt{mgcv} (ridge) and \texttt{shrinkage} (Bay\_glo) from \emph{all} 400 subsets: 2.5\%, 50\% (dot), 97.5\% quantiles. Right panel (grey):
posterior estimates of ridge penalty for 10 random subsets using Bay\_glo}\label{varpen}
\end{figure}

Uncertainty of the predictions $\boldeta$ results from the variability of $\bbeta$, which includes two components: one conditional on fixed penalty parameter(s) $\blambda$ and one reflecting the variability of $\blambda$. \cite{riley2021penalization} show that the latter, the ridge penalties, may be very variable across small subsets. Indeed, the left-hand side of Figure \ref{varpen} confirms this for our data: the penalty estimates may differ 2-3 natural logs in magnitude from one subset to another ($n=50$), with standard ridge and Bay\_glo alike. Hence, the extra regularization of Bay\_glo has little effect here. When multiple penalties are estimated, however, extra regularization can improve stability of those, as illustrated by Supp. Fig. \ref{varpen2}: Bay\_2 compares favourably to ridge\_2 for stability of $\lambda_2$. Nevertheless, between subset variability of $\blambda$ remains considerably large. Therefore, we argue that it is important to propagate the uncertainty of $\blambda$ into estimation of $\bbeta$ when analysing \emph{one} subset, provided that this reflects the \emph{between} subset variability, as the latter is usually not available. The right-hand side of Figure \ref{varpen} shows this: when using Bay\_glo the posterior variability of $\blambda$ for ten random subsets approximates the between subset variability of the ridge penalty (left-hand side) fairly well, in particular in order of magnitude. The penalties are usually not the primary estimands of interest; what matters most is to what extent their variability propagates towards $\bbeta$ and, eventually, the predictions $\boldeta$. Therefore, we finalize the comparison by evaluating coverages of the confidence intervals of $\boldeta$.

\para
As before, we focus on OLS, ridge and ridge\_2 and contrast these with Bay\_loc, Bay\_glo and Bay\_2 in Table \ref{covtable}. Both ridge and ridge\_2 are used with uncertainty propagation of the penalties, as implemented in \texttt{mgcv}'s \texttt{predict.gam} function with argument \texttt{unconditional = TRUE}. First, while we observe that OLS and Bay\_loc are competitive in terms of coverage, the latter results in narrower, and hence slightly more useful, intervals. Second, unlike for the introductory example, ridge and Bay\_glo are competitive in terms of coverage, and also in terms of mean width of the intervals. Finally, for the grouped penalties the extra shrinkage imposed by Bay\_2 pays off as it maintains good coverage, unlike ridge\_2, which seems to render too narrow intervals. Note that in the latter case, the 10\% quantile of the coverage of the predictions (as computed from the test individuals) drops to only 0.772 for ridge\_2, whereas it equals 0.925 for Bay\_2.
Note that the mean coverages of ridge and ridge\_2 drop slightly (from 0.944 to 0.920 and from 0.896 to 0.864, respectively) when not adjusting the standard errors (\texttt{unconditional = FALSE} in \texttt{predict.gam}). For $n=200$, results are qualitatively similar, although the intervals of OLS are now somewhat narrower than those of Bay\_loc (Supp Table \ref{covtablen200}).


\begin{table}[ht]
\centering
\begin{tabular}{|c|rr||rr|}
  \hline
   & \multicolumn{2}{c||}{Coverage} & \multicolumn{2}{c|}{Width}\\
Methods  & Classical & Bayes & Classical & Bayes \\\hline
OLS, Bay\_loc   & 0.939 & 0.971 & 2.334 & 2.149 \\
ridge, Bay\_glo  & 0.944 & 0.961 & 1.521 & 1.607 \\
ridge\_2,  Bay\_2  & 0.896 & 0.966 & 1.216 & 1.470 \\
   \hline
\end{tabular}
\caption{Mean coverage (target: 0.95) and width of confidence intervals of predictions for $n=50$, contrasting classical method with Bayesian counterpart}\label{covtable}
\end{table}

\subsection{Conclusions, linear case}
Below we list the conclusions from the linear regression case.
\begin{enumerate}
  \item Shrinkage benefits prediction accuracy compared to OLS and stepwise selection.
  \item Differential shrinkage for groups of covariates may substantially benefit prediction accuracy compared to global shrinkage. Therefore, the same prediction accuracy can be achieved with a smaller sample size.
  \item Penalty estimates vary substantially between small subsets of samples. Hence, it is relevant to propagate this variability when quantifying uncertainty of the predictions.
  \item Shrinkage may indeed lead to some level of miscalibration, although minimally for Bayesian shrinkage with a local C$^+$ prior. The latter is competitive to OLS on this matter and on coverage, while rendering better prediction accuracy. Prediction accuracy is worse, though, than that of the global regularization methods.
  \item Bayesian shrinkage with a grouped C$^+$ prior outperforms its frequentist counterparts in terms of interval estimation. In case of global shrinkage, the two are very competitive, with the latter slightly less conservative.
\end{enumerate}

%

\section{Calibration for logistic regression: external simulation}
Here, we study the small sample logistic regression setting. This differs from the linear one in three essential ways: a) binary outcomes are much less information-rich than continuous ones; b) the logistic function flattens the impact of large $\beta$'s; and c) the maximum likelihood estimator (MLE) is biased \cite[]{firth1993bias}.
We focus on prediction accuracy and calibration, as standard shrinkage methods like lasso and ridge with cross-validated penalties generally improve the first, but may deteriorate the latter. This is intrinsic due to the introduced bias, but may be worsened by the conventional tuning of the penalty parameter: minimization of cross-validated prediction error, which may be far from optimal for calibration. Hence, we discuss some alternatives. Firth's solution to adjust bias \cite[]{firth1993bias} uses a fixed penalty, defined on the level of the information matrix, and was shown to perform favourably to the tuning methods \cite[]{van2020regression}.
\cite{sullivan2013bayesian} propose a similar, alternative solution, which also accounts for the aforementioned issues a) and b): simply fix the prior variance of the $\beta$'s to 0.5, equivalent to setting the ridge penalty $\lambda = 1/0.5=2$. Their argument is that for the small sample logistic setting one has to `dare' to be fairly informative, as the outcomes are information-poor. Based on this, we propose a somewhat more objective alternative that does not require any tuning either: use Bay\_glo or Bay\_loc as defined in Table \ref{priors}, but with a $C^+(0,\sqrt{0.5})$ instead of a $C^+(0,1)$ prior for the standard deviation, as the former matches to a median prior variance of 0.5. These priors are depicted in Supp. Fig. \ref{halfcauchy}. So, we allow for some variation in the prior variances, thereby enabling adaptation to cases in which (some of the) $\beta$'s are more (or less) extreme.

\begin{figure}[h]
\centering
\includegraphics[scale=0.45]{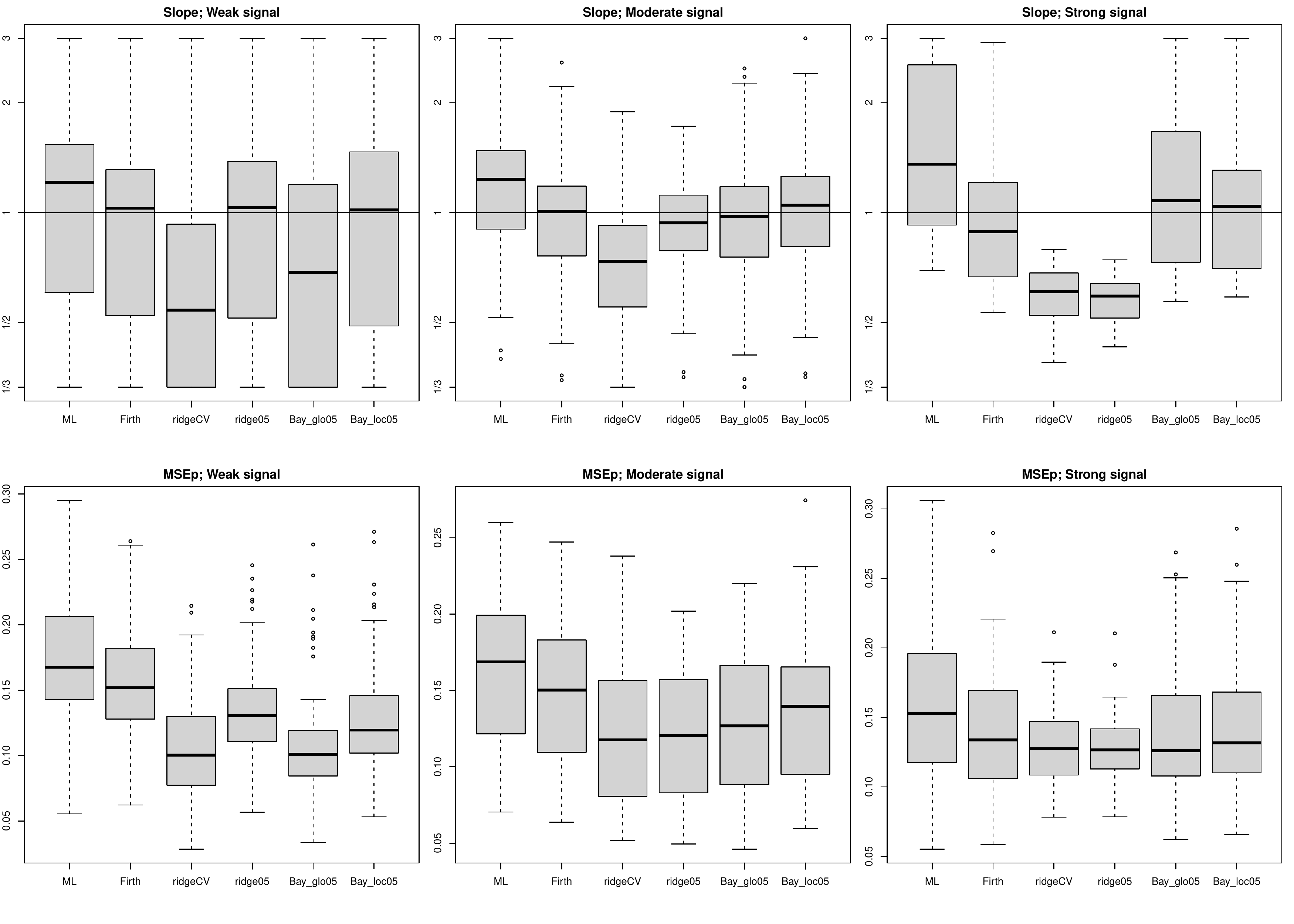}
\caption{Calibration (top; cslope: winsorized at $(1/3,3)$, log-scale) and prediction accuracy (bottom; MSEp) for logistic regression simulations}\label{logistic}
\end{figure}

\para
For evaluation we follow one of the simulation set-ups by \cite{van2020regression}: $n=50, \bbeta = (\beta_1, \ldots, \beta_5)= (0.2,0.2,0.2,0.5,0.8)$ and intercept $\beta_0$ tuned such that the average event probability equals 0.5, rendering a challenging setting of 25/5 = 5 events per covariate. Covariates $X$ are simulated from a multivariate normal with means 0, variances 1, and correlations 0.5. Response $Y$ is then simulated from a Bernoulli with success probabilities $1/(1+\exp(-\beta_0 - X\bbeta)).$  We refer to this setting as `Moderate signal'. Additionally, to study how results adapt to the signal strength, we simulate the `Weak and Strong signal' settings by using $\bbeta_{\text{weak}} = \bbeta/3$, and  $\bbeta_{\text{strong}} = 3\bbeta$.
Calibration is assessed by the `cslope', which results from regressing the estimated linear predictors against the true ones (using the real $\bbeta$) for a large test set. A slope close to 1 indicates good calibration. Note that \cite{van2020regression} report the classical calibration slope, which results from regressing observations on predictions. The latter measures in fact a mix of calibration and spread of the predictions \cite[]{stevens2020validation}, which is why we opt for the cslope.
In our case, too strong shrinkage will render cslope $\ll 1$.
The MSE of the predictions (MSEp) is assessed by the average squared error of those predicted probabilities w.r.t. the true ones.
Figure \ref{logistic} displays the results based on 50 training sets for ML (maximum likelihood); Firth; ridgeCV: ridge with cross-validated penalty; ridge05: ridge with fixed penalty 1/0.5; Bay\_glo05 and Bay\_loc05: global and local regularization with $C^+(0,\sqrt{0.5})$ priors for the standard deviation(s). Below we list the conclusions.

\para
\begin{itemize}
  \item We confirm that ridgeCV performs well in terms of prediction accuracy, but poor on calibration, with clopes $\ll 1$. Firth calibrates better than both ML and ridgeCV.
  \item The solution by \cite{sullivan2013bayesian}, ridge05, is competitive to Firth in terms of calibration, when the signal is weak or moderate,
   but overshrinking when the signal is strong. It is generally superior to ML and Firth for prediction.
  \item Bay\_glo05 is generally fairly competitive to ridgeCV in terms of prediction accuracy. On average it calibrates better, but shows more variability. It calibrates less well than Firth, and Bay\_loc05.
  \item Bay\_loc05 is competitive to Firth in terms of calibration, but is superior for prediction in particular in the weak signal setting.
  \item Bay\_loc05 is competitive to ridge05 in terms of calibration. It adapts better to the change in signal at the price of somewhat more variation. It is competitive in terms of prediction accuracy, taking into account that variation towards low MSEp is actually desirable.
\end{itemize}
Supp Fig \ref{logisticn100} and \ref{logisticaddzeros} present the results of two alternative scenarios (larger sample size: $n=100$, and adding five $\beta$'s equalling 0). Results are qualitatively similar to those presented here. For completeness, we have also co-plotted the (inverted) cslope and the classical calibration slope for the original simulation scenario in Supp Fig \ref{cslopecalslope}. These two agree fairly well for the strong signal setting, but less so for the weak and moderate signal settings, as the classical calibration slopes are more affected by the variability of the predictions.

\para
All-in-all, we conclude that for calibration Firth, ridge05, and Bay\_loc05 are competitive to one another and superior to ML and the global penalization methods with adaptive penalty, ridgeCV and Bay\_glo05. The latter two, though, are superior in terms of prediction accuracy when the signal is weak. In that case Bay\_loc05 outperforms Firth in terms of prediction accuracy. Unlike ridge05, Bay\_loc05 adapts to the signal strength, but shows somewhat more variability due to the non-fixed penalty. Note that an important advantage of Bay\_loc05 is that unlike Firth and ridge05, it straightforwardly provides uncertainty estimation of the predictions, including the uncertainties of the penalties.

\section{Software and reproducibility}
\subsection{Software}
Below, we list the software packages used for the various models.
\begin{enumerate}
  \item OLS and stepwise selection: \texttt{R}'s \texttt{lm} and \texttt{step} functions.
  \item Linear ridge, global and grouped penalty: \texttt{gam} function in \texttt{R}-package \texttt{mgcv} \cite[]{wood2011fast}. Unlike \texttt{glmnet} this allows for uncertainty computations.
  \item Logistic ridge: \texttt{glmnet}, as this was also used by \cite{van2020regression}. Firth's correction: \texttt{logistf} \cite[]{heinze2002solution}.
  \item Lasso: \texttt{glmnet} \cite[]{Friedman2010}.
  \item Bayes, linear: \texttt{shrinkage} software \cite[]{rshrinkage}, which is optimized for computational efficiency, hence convenient for the repeated calculations on data splits. Results highly agree with those of \texttt{r-stan}.
  \item Bayes, logistic: \texttt{r-stan} \cite[]{rstan}.
\end{enumerate}

\subsection{Data sharing}
As is the case for many cohort studies, our primary data source, the Helius data, cannot be shared publicly. This is impeding methodological studies like this, because such large $N$ studies are very useful to check results of various methods on small subsets as we did. Therefore, we share a synthetic copy of our data set which may be used to a) qualitatively reproduce our results; or b) evaluate other methods than those proposed here. We verified whether results from the synthetic copy agreed with those from the real data, which is the case (see Supp Fig \ref{pmserealsynth} for MSEp). The imputation-based method to generate the synthetic data is described in the Supp Mat.

\subsection{Do it yourself}
One may certainly argue that results and conclusions might differ somewhat for other data than those used here.
The following scheme may be useful to assess what shrinkage method works best for the data at hand. Prediction accuracy can be assessed by cross-validation principles. To evaluate calibration and coverage, one needs to know `real' predictions in a setting that mimics one's own. For that, run an OLS or, preferrably, the more stable Bay\_loc on the data and obtain $\hat{\bbeta}$ as well as an estimate of the noise. Then simulate response $y$ many times from the regression model using the real design matrix $X$ and those estimated parameters. Then, our scripts (or other software for the proposed models) can be used to assess the performances of the methods in one's own data-centric simulation setting.

\subsection{Availability}
Annotated software scripts and \texttt{R}-markdown files to reproduce our results are available from github: \url{https://github.com/markvdwiel/ThinkBeforeShrink}.
This repository also contains a synthetic copy of the Helius data and an example script to guide users.

\section{Discussion}
We observed that grouping of covariates to allow for differential penalties may improve predictive performance. Here, we only studied the low-dimensional linear regression setting, but we have observed similar results for the high-dimensional logistic ridge setting \cite[]{WielGRridge, van2021flexible}. One could argue that the grouping of covariates is subjective. However, we demonstrated that the results are robust against misspecification of the groups.  Moreover, it is much less subjective than a commonly used strategy, as stated in \cite{van2020regression}: ``Alternatively, a less complicated model can be considered, for example by discarding many predictors a priori.''. The latter strategy basically assigns an infinite penalty to the excluded covariates a priori.  We believe it is better to leave those in, but allow differential shrinkage for those as compared to the others.

\para
If calibration is key, global shrinkage methods with flexible penalties, both frequentist and Bayesian ones, are usually not very suitable. This is not surprising as these penalty parameter(s) adapt to predict optimally in terms of accuracy, which is a different goal than calibration. We showed that Bayesian local shrinkage is a good alternative which competes with conventional methods that do not tune penalties, such as OLS, ML, Firth and ridge with a fixed penalty. Note that the importance of having calibrated predictions depends on a) the possibility of recalibrating predictions in a later stage of the study; and b) the need or desire to interpret the predictions on an absolute scale. For the latter, the design of the study is also very relevant; often recalibration will be needed anyhow when applying the predictor to a population with somewhat different characteristics than the one represented by the study, a very common case for medical studies.

\para
While the grouping of covariates can aid in improving prediction accuracy, it may deteriorate coverage of the confidence intervals of predictions in the classical setting. Here, the additional regularization by the $C^+(0,1)$ prior as invoked by the Bayesian procedure helps to improve coverage, while maintaining competitiveness on prediction accuracy and calibration. The importance of quantifying uncertainty of the predictions differs from study to study. However, in any study it will be useful to know it, also to assess whether the sample size should be increased to lower this uncertainty to an acceptable level.

\para
We discussed the linear and logistic case, which could be regarded as two extremes: the former being information-rich with unbiased ML (= OLS) estimates, the latter being information-poor with biased ML estimates. This means that in the latter case mild shrinkage is also beneficial for calibration. Moreover, it may be relatively more beneficial to a use a subjective prior (or penalty) in the logistic setting. We discussed two examples that performed well, prior variance equal to 1/2 \cite[]{sullivan2013bayesian}, or less subjective, a hyperprior for the variance with median equal to 1/2. A good alternative may be to use a historical prior, which is tuned to available data from similar studies \cite[]{Neuenschwander2010,maclehose2014applications}. Then, the solution with a hyperprior may be preferable over one that fixes the variance, as the former allows the current study to deviate somewhat more when it would not behave similarly.

\para
For our applications, computational time was not an issue at all: all methods ran within seconds for given data sets. Nevertheless, the classical methods fitted substantially faster than the Bayesian ones, rendering the former an edge for analysing larger scale studies. Whether this balances against the demonstrated benefit of higher-level shrinkage that Bayesian methods can provide, will depend on the main aim(s) of the study (prediction accuracy / calibration / uncertainty quantification). In general, Bayesian methods gain popularity in epidemiological research \cite{maclehose2014applications}, which assists their acceptance for general use. Note that, on purpose, we did not tweak the Bayesian methods in terms of convergence checks or use of other hyper-priors, as these would imply further tuning, rendering the comparison with less flexible methods unfair. In terms of software, we recommend to use either \texttt{mgcv}, or \texttt{R-stan} with the $C^+$ priors, as both incorporate estimation of multiple penalty parameters and provide intervals with, in most settings, fairly good coverage. The latter has an edge when calibration is key as it allows for local penalties, or when stabilization of grouped penalties is relevant, e.g. for uncertainty quantification. Supp Fig \ref{flowchart} shows a flowchart with our recommendations on when to use which method.

\para
Our work is limited in scope, so we discuss several important extensions. The first one is variable selection. As this is a different goal than prediction, other priors and penalties should be discussed.
In the classical setting many variants of the lasso (adaptive, group, hierarchical) become relevant, compared to more traditional stepwise selection techniques. In the Bayesian setting, one may wish to include spike-and-slab priors and/or Zellner's g-prior, as these are targeting for variable selection \cite[]{george2000calibration}. Alternatively, posterior selection techniques that apply to the fairly dense ridge-type methods used here may be very competitive to those more sparse formulations \cite[]{Bondell2012}.

\para
A second extension is to study the effect of shrinkage in the context of causal inference. Many causal inference frameworks make use of prediction methods to account for confounding or non-random treatment allocation. Unlike in predictive modelling, the emphasis in causal inference is primarily placed on bias, and much less so on variance reduction. This makes the use of shrinkage less natural, in particular for the treatment effect. Nevertheless, some (local) shrinkage of the other covariates may be beneficial
when $p$ is relatively large compared to $n$, possibly in combination with double-robust estimation \cite[]{avagyan2021high} to counter misspecification of the model.

\para
A third extension is the multi-regression setting, as often encountered in high-dimensional multiple testing setting, e.g. when relating gene expression to phenotype, correcting for confounders like age and gender. In such a setting, shrinking effects across similar features (e.g. genes) may improve effect size estimates and multiple testing properties \cite[]{WielShrinkSeq}. Moreover, a multi-regression setting allows tuning of hyper-parameters of the penalty's prior across features using empirical Bayes \cite[]{Leday2017}.


\para
Shrinkage is only a partial solution for underpowered studies. In the end, increase of sample size will often be needed to draw firm conclusions \cite[]{riley2020calculating}. However, we do believe shrinkage may play an important role in the research cycle, which for many practical reasons often starts with a fairly small study.  Then, as demonstrated, well-thought global or grouped shrinkage methods can help to better assess the predictive potential of the study and quantify uncertainty of the predictions, whereas local shrinkage methods are able to reduce bias and improve calibration. Therefore, these shrinkage methods aid in deciding whether and how to extend the study.

\bibliographystyle{C://Synchr///Stylefiles/author_short3} 
\bibliography{C://Synchr//Bibfiles//bibarrays, C://Synchr//Bibfiles//synth}      

\section{Figures}


%
%
%

\section{Acknowledgements}
The Amsterdam University Medical Centres and the Public Health Service of Amsterdam (GGD Amsterdam) provided core financial support for HELIUS. The HELIUS study is also funded by research grants of the Dutch Heart Foundation (Hartstichting; grant no. 2010T084), the Netherlands Organization for Health Research and Development (ZonMw; grant no. 200500003), the European Integration Fund (EIF; grant no. 2013EIF013) and the European Union (Seventh Framework Programme, FP-7; grant no. 278901). Moreover, we thank Hans Berkhof for discussions on several aspects of this manuscript.
\clearpage

\section{Supplement, incl. Figures}

\subsection{Synthetic data}
Data from the Helius study \cite[]{stronks_unravelling_2013} were used as a running example throughout the manuscript, but they cannot be shared due to privacy regulations. To enhance reproducibility of our results, we provide a synthetic copy of the Helius study data to render qualitatively similar results as compared to the real data. This supplement describes construction of the synthetic data set.

In essence, the problem was approached from a missing data perspective. After applying standardization and the removal of rows with missing data (as described in the main manuscript), the original data set of size $n \times m$ was augmented with an empty data set of size $n \times m$. Subsequently, the empty rows were imputed using imputation by chained equations \cite[]{white_multiple_2011, buuren_flexible_2018} as implemented in the \texttt{mice} package \cite[]{buuren_mice_2011} in \texttt{R}. This procedure makes use of univariate imputation models for every incomplete variable, and uses an iterative updating scheme that iterates between imputation and updating of the imputations models. The specific application of (multiple) imputation by chained equations for the creation of synthetic data was described by \cite{volker_anonymiced_2021}.

For this particular application, the  variables systolic blood pressure, age, gender, BMI, ethnicity (5 levels), smoking (binary), packyears, coffee (binary), glucose (log), and cholesterol, had to be imputed. Predictive mean matching \cite[]{buuren_flexible_2018} was used for all continuous variables, logistic regression imputation was used for binary variables, and a multinomial imputation model was used to impute ethnicity. All imputation models were linear additive models conditional on all of the remaining variables. For example, the imputation models for age was a linear additive model with main effects of systolic blood pressure, age, gender, BMI, ethnicity (dummy coded, 4 df), smoking (binary), packyears, coffee (binary), log glucose, and cholesterol level. A single imputed data set was created to represent a synthetic version of the helius data. Traceplots were stable after the chosen number of 25 iterations for the \textit{mice} algorithm. The synthetic data set was found to closely resemble the original data with respect to mean structure, variance, and covariance structure, with a mean absolute deviation from the original data $< 0.01$ on the standardized data for all three. The synthetic data set did not contain any duplicated rows from the original data. As a further check, a random forest was fitted with the aim to distinguish the original and synthetic data \cite[]{breiman_random_2001, liaw_classification_2002}. Half of the data was used for training, and half for testing, and 500 trees were fitted. The test Brier score was 0.23 (with random providing 0.25), indicating that the forest could not accurately separate real and synthetic data.

It is worth noting that the linear additive nature of the imputation models suited our purposes well, but that it does not provide a general solution for synthetic data generations, because it does not reflect possible non-linear relations and interactions. Other possibilities are available, such as multiple imputation using classification regression trees \cite[]{volker_anonymiced_2021}. Also, it should be noted that use of a single imputed data set does not reflect the uncertainty of the imputed (synthetic) values. For synthetic data to be used to replicate inference results, where this uncertainty is key, multiple imputed/synthetic sets would be required \cite[]{volker_anonymiced_2021}.

\subsection{Figures}
\setcounter{figure}{0}
\begin{figure}[h]
\centering
\includegraphics[scale=0.9]{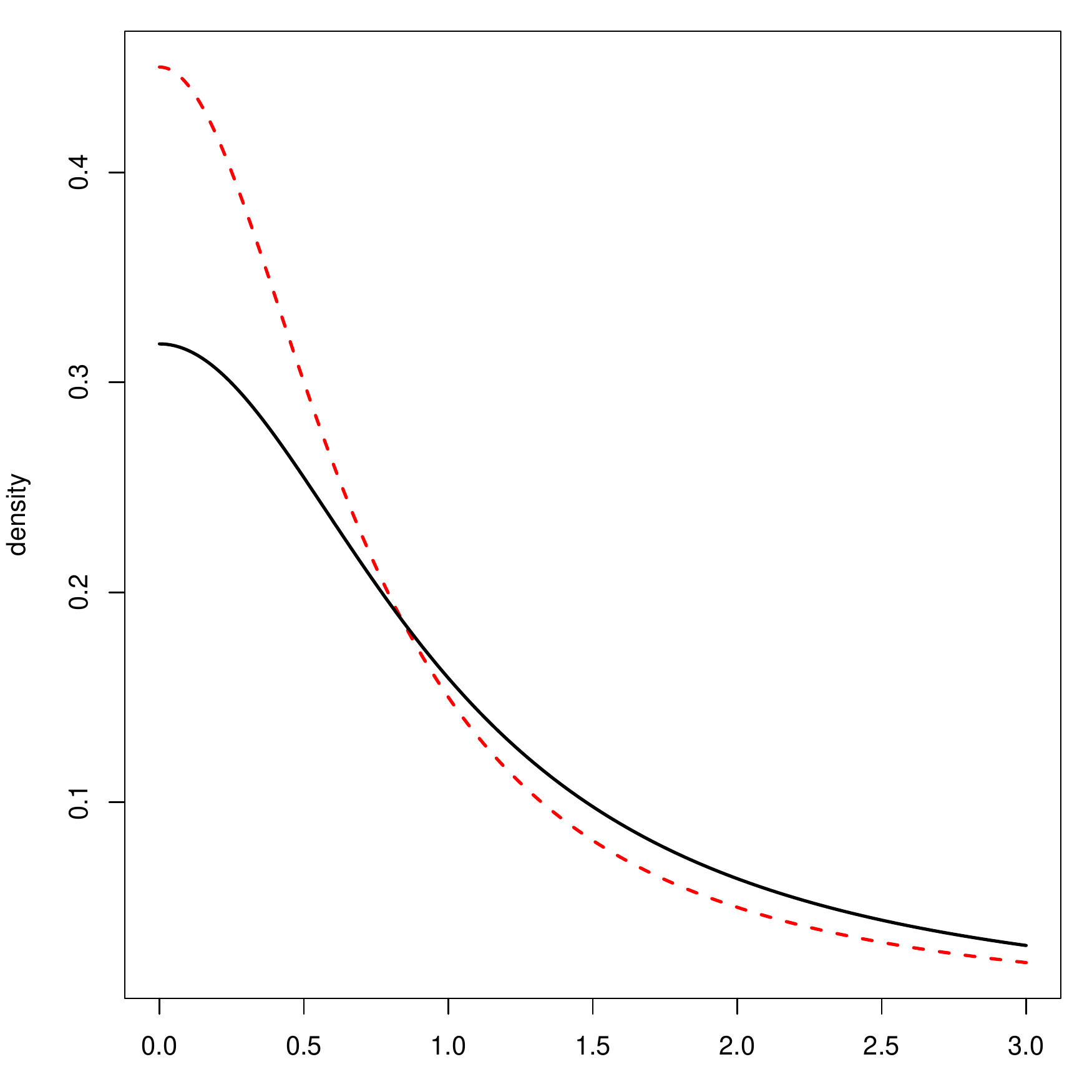}
\caption{Half-Cauchy priors: $C^+(0,1)$ (solid), $C^+(0,\sqrt{0.5})$ (dashed)}\label{halfcauchy}
\end{figure}

\begin{figure}[h]
\centering
\includegraphics[scale=0.9]{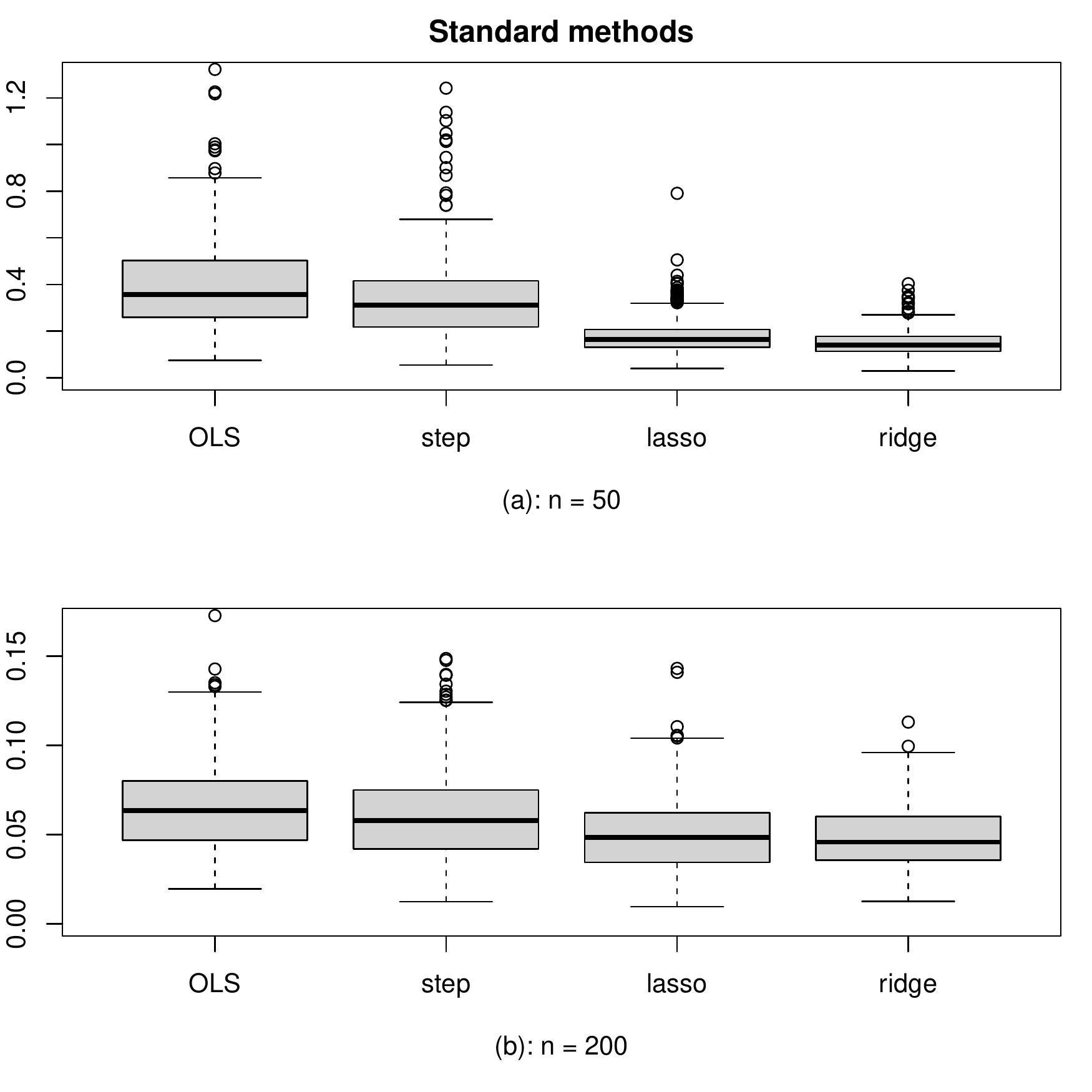}
\caption{MSEp (y-axis) for several standard methods across 400 subsets}\label{standard}
\end{figure}

%
%
\begin{figure}[h]
\centering
\includegraphics[scale=0.9]{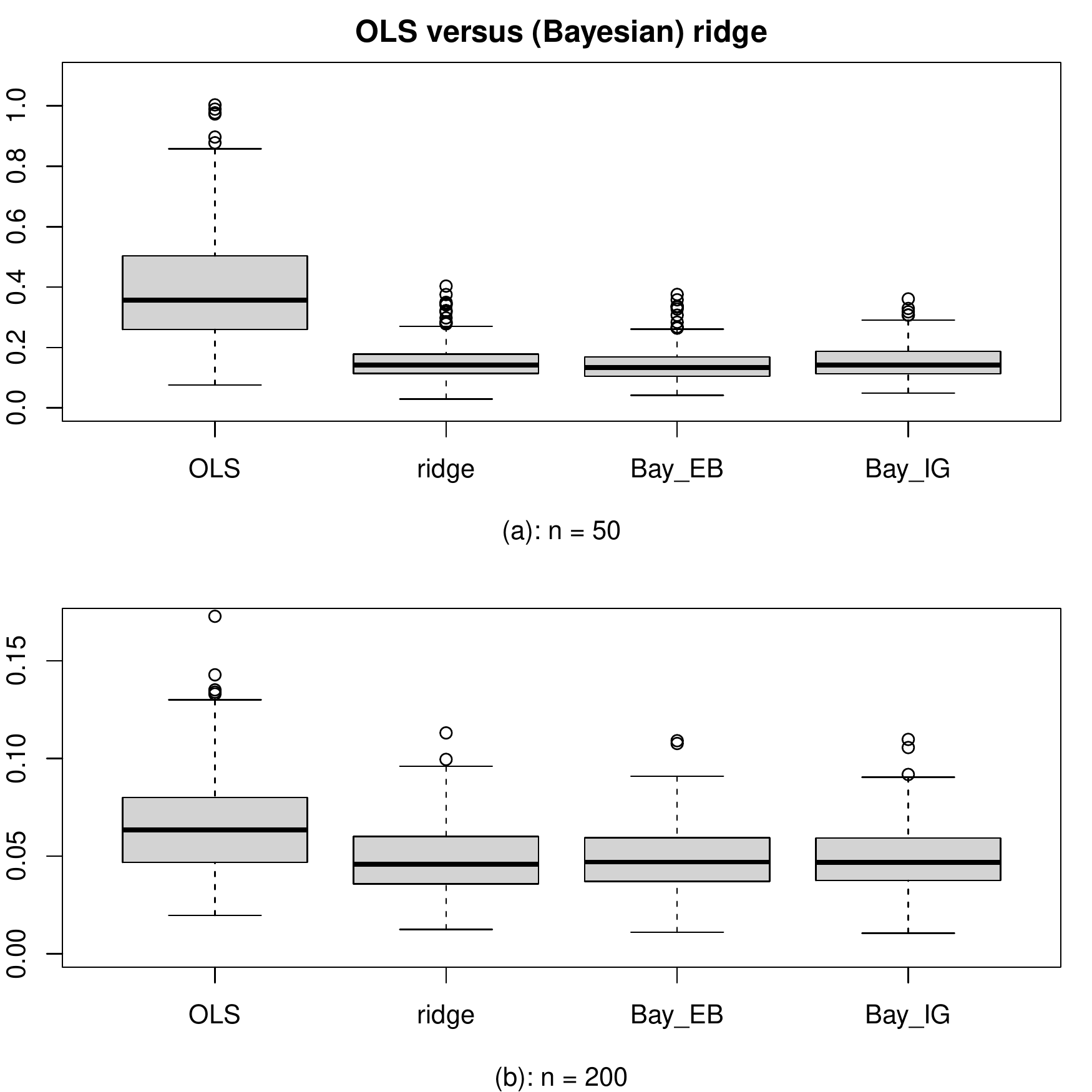}
\caption{MSEp (y-axis) for standard ridge variants across 400 subsets}\label{ridgevariations}
\end{figure}

\begin{figure}[h]
\centering
\includegraphics[scale=0.9]{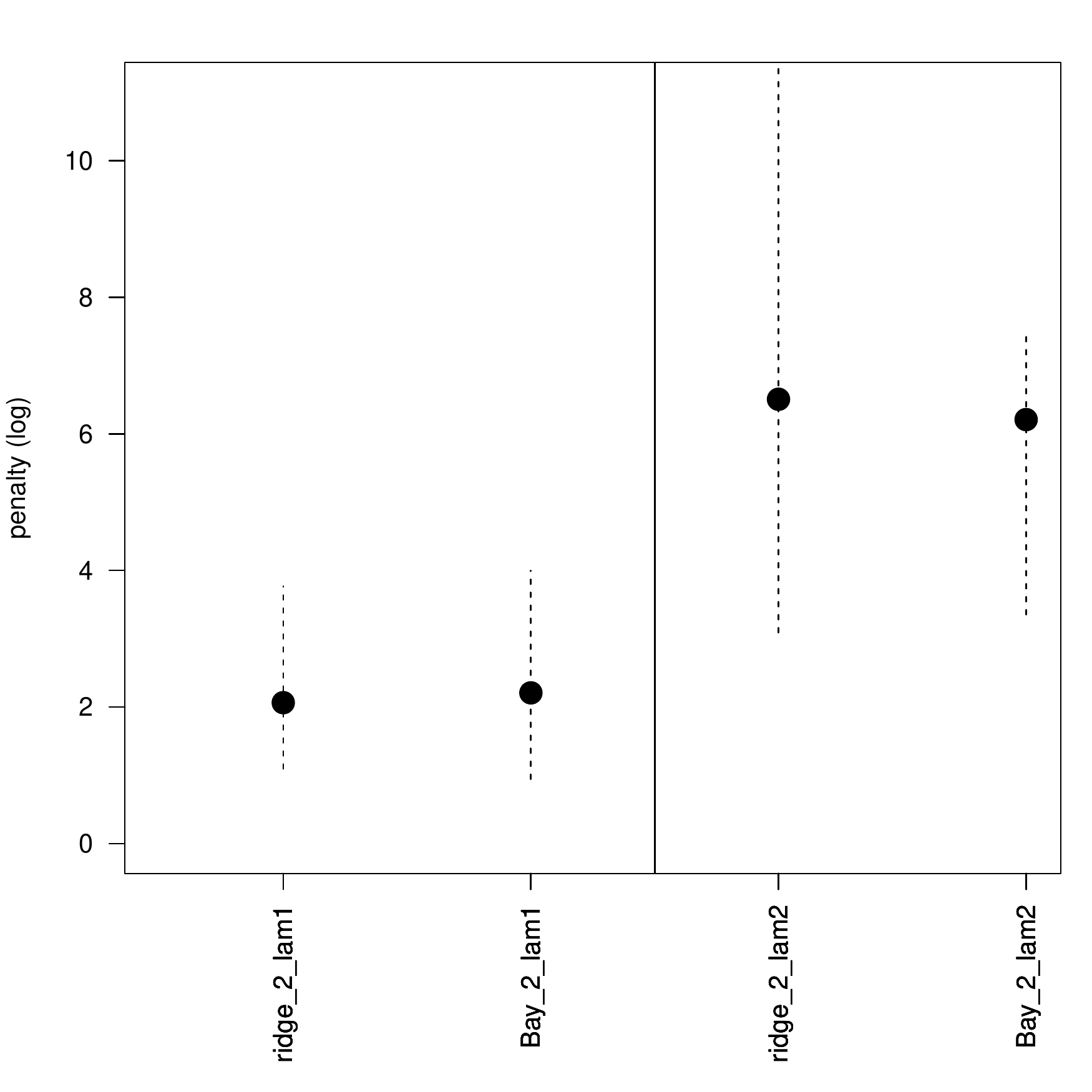}
\caption{Variability of multi-group (2) ridge penalties. Y-axis: ridge penalties on natural log-scale. Ridge penalties (lam1, lam2) estimated by
\texttt{mgcv} (ridge\_2) and \texttt{shrinkage} (Bay\_2) from all 400 subsets: 2.5\%, 50\% (dot), 97.5\% quantiles}\label{varpen2}
\end{figure}


%

\begin{figure}[h]
\centering
\includegraphics[scale=0.75]{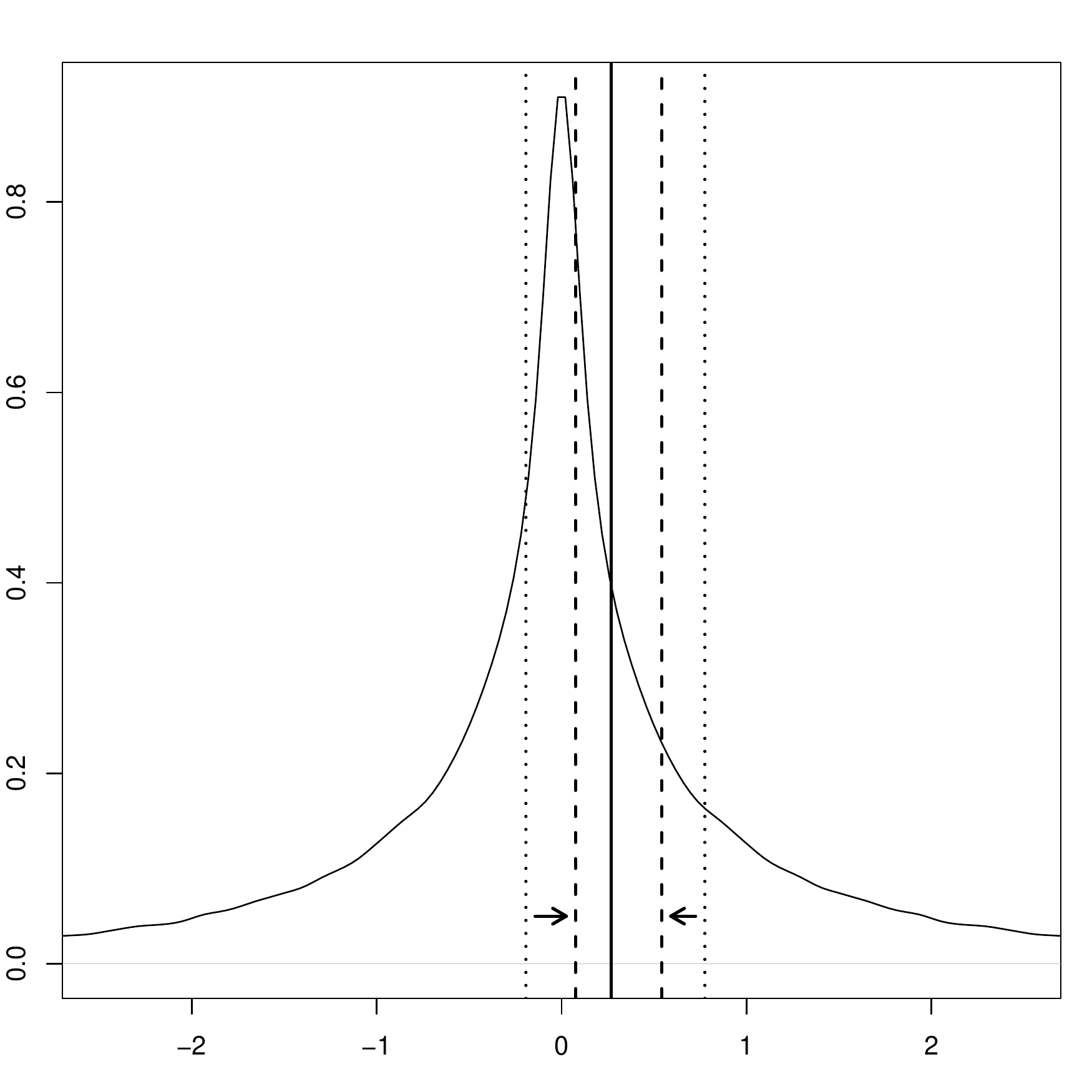}
\caption{Normal-$C^{+}(0,1)$ prior and estimates of $\beta_{\text{BMI}}$ for two subsets of size $n=50$. Vertical lines: true value of $\beta_{\text{BMI}}$ (solid), OLS estimates (dotted) and Bay\_loc estimates (dashed). Arrows indicate direction of shrinkage by Bay\_loc.}\label{priorbeta}
\end{figure}

\begin{figure}[h]
\centering
\includegraphics[scale=0.43]{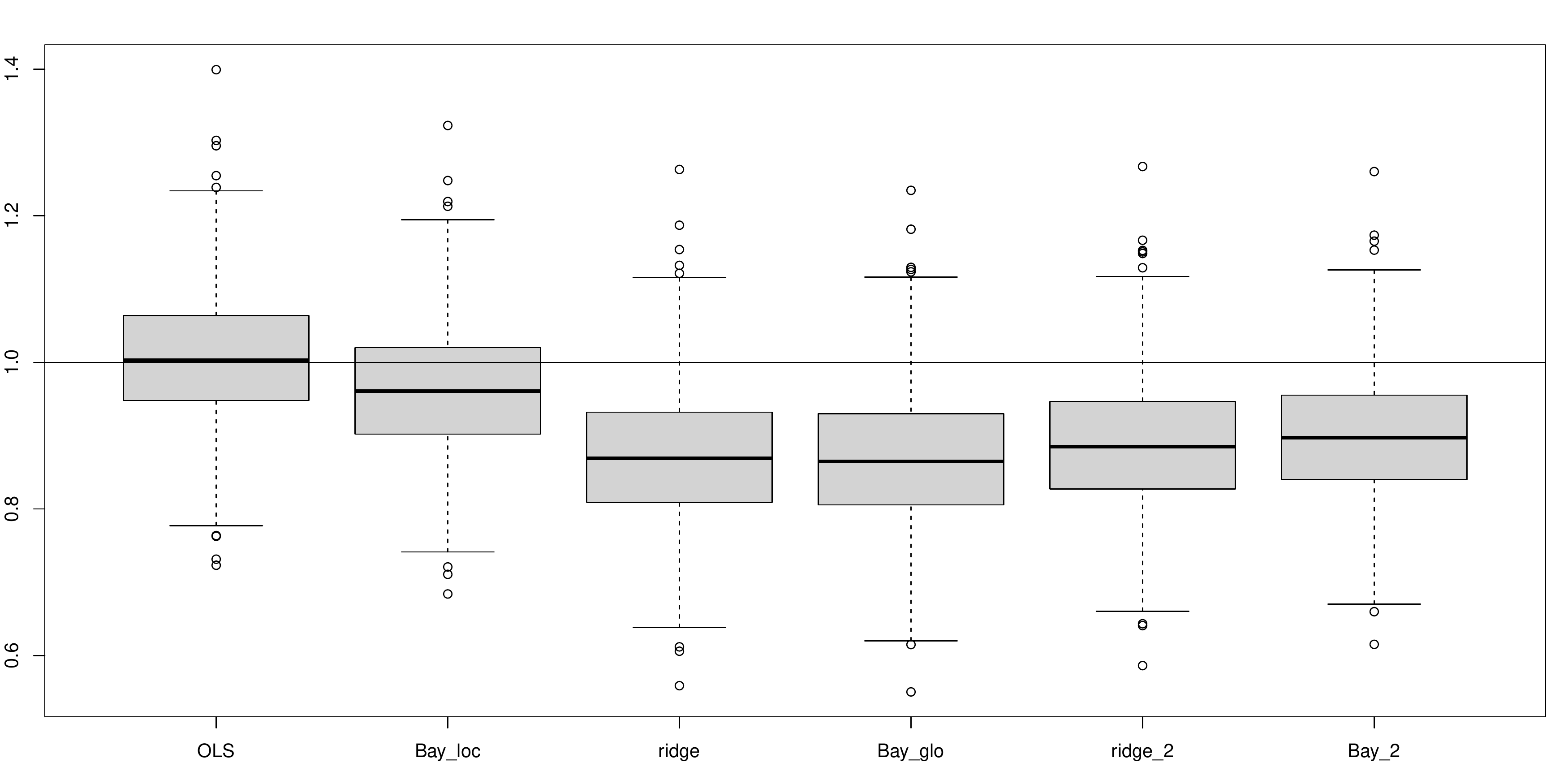}
\caption{Calibration: cslopes across all 400 subsets of size $n=200$.}\label{calibrationn200}
\end{figure}

\begin{figure}[h]
\centering
\includegraphics[scale=0.75]{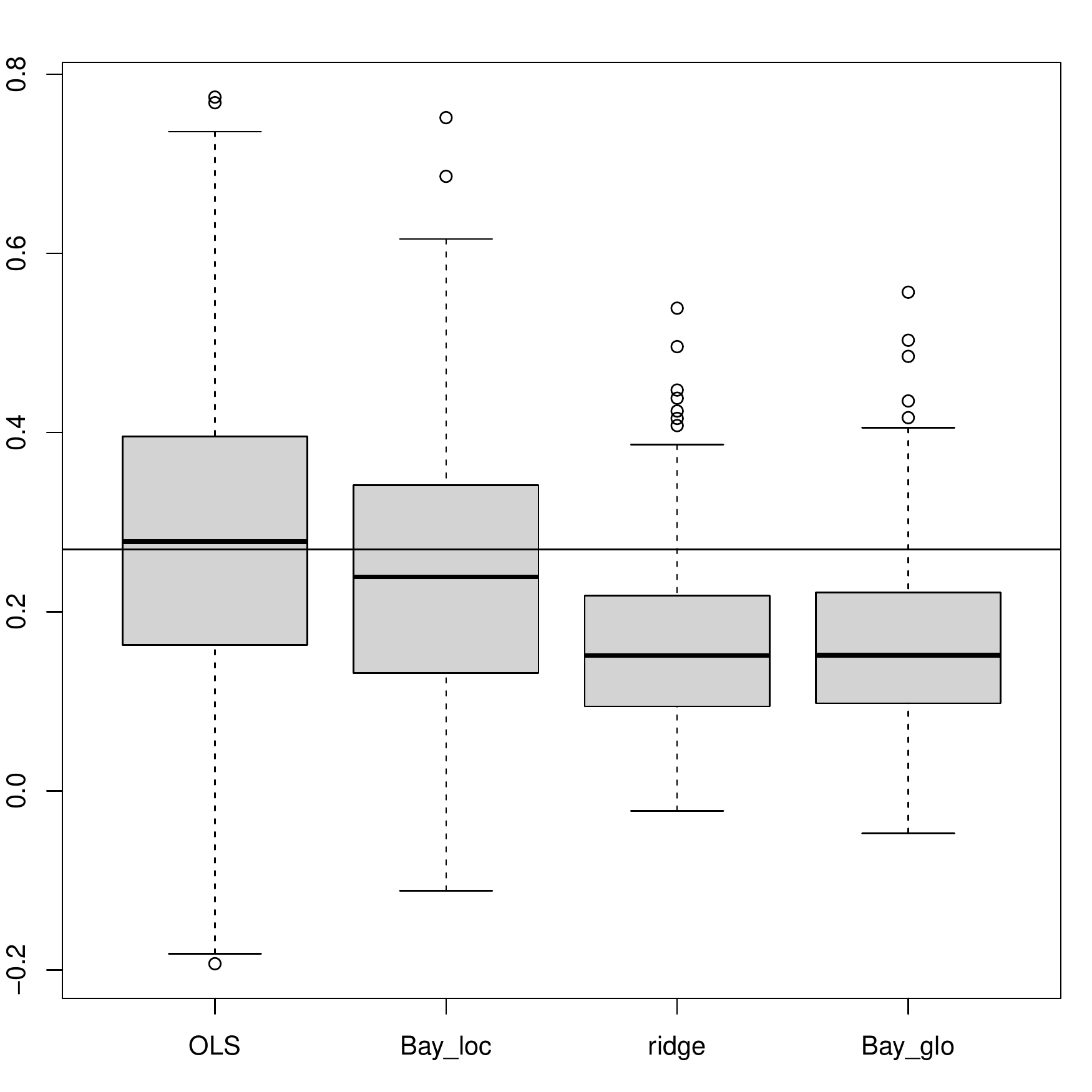}
\caption{Estimates of $\beta_{\text{BMI}}$ for subsets of size $n=50$. Horizontal line: true value of $\beta_{\text{BMI}}$. }\label{bmicoefs}
\end{figure}

\begin{table}[ht]
\centering
\begin{tabular}{|c|rr||rr|}
  \hline
   & \multicolumn{2}{c||}{Coverage} & \multicolumn{2}{c|}{Width}\\
Methods           & Classical & Bayes & Classical & Bayes \\\hline
OLS, Bay\_loc     & 0.949 & 0.984 & 0.980 & 1.186 \\
ridge, Bay\_glo    & 0.953 & 0.973 & 0.864 & 1.000 \\
ridge\_2,  Bay\_2  & 0.905 & 0.952 & 0.746 & 0.870 \\
   \hline
\end{tabular}
\caption{Mean coverage (target: 0.95) and width of confidence intervals of predictions for $n=200$}\label{covtablen200}
\end{table}

\begin{figure}[h]
\centering
\includegraphics[scale=0.45]{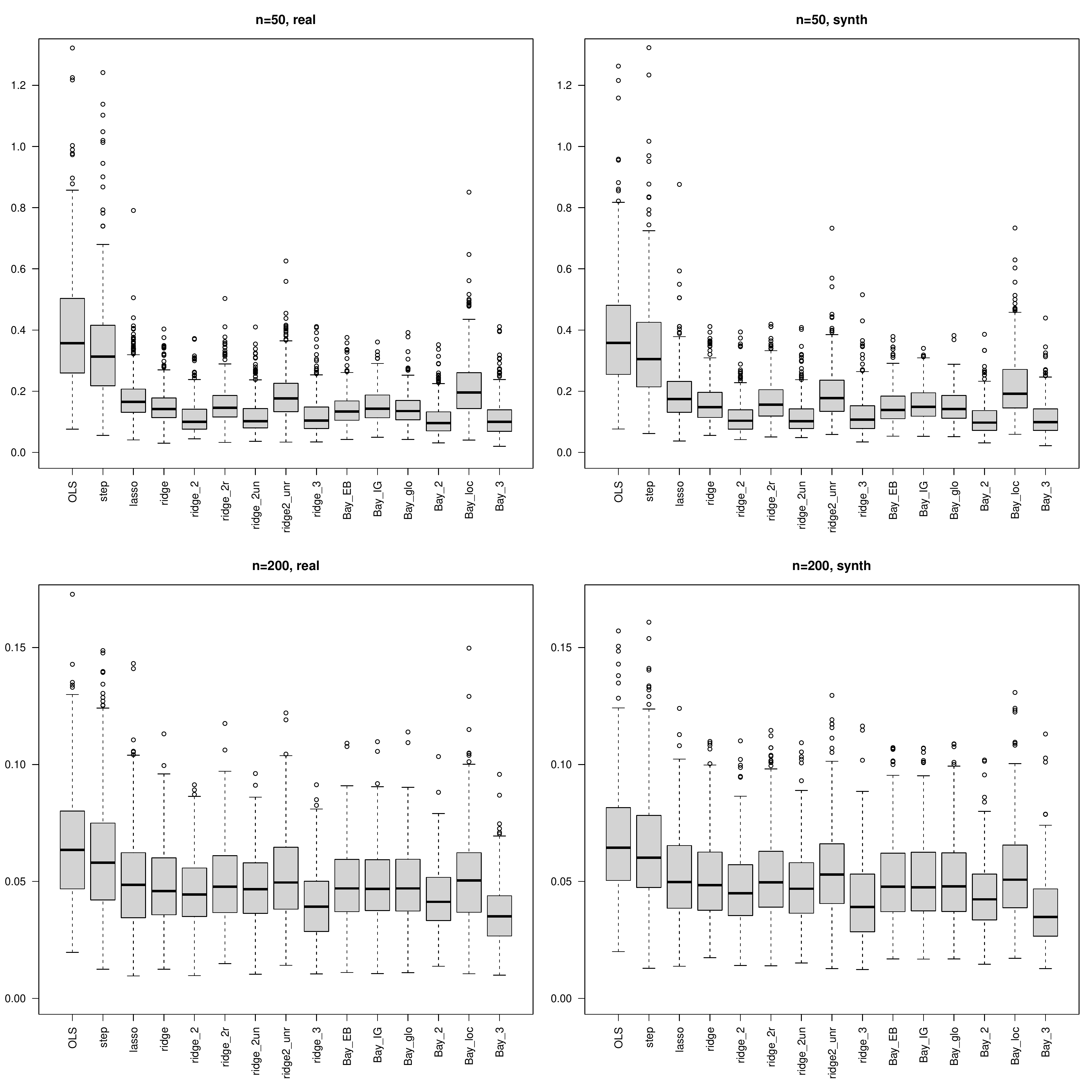}
\caption{MSE of the predictions for real (left) and synthetic (right) data. Upper row: $n=50$, bottom row: $n=200$}\label{pmserealsynth}
\end{figure}

\begin{figure}[h]
\centering
\includegraphics[scale=0.5]{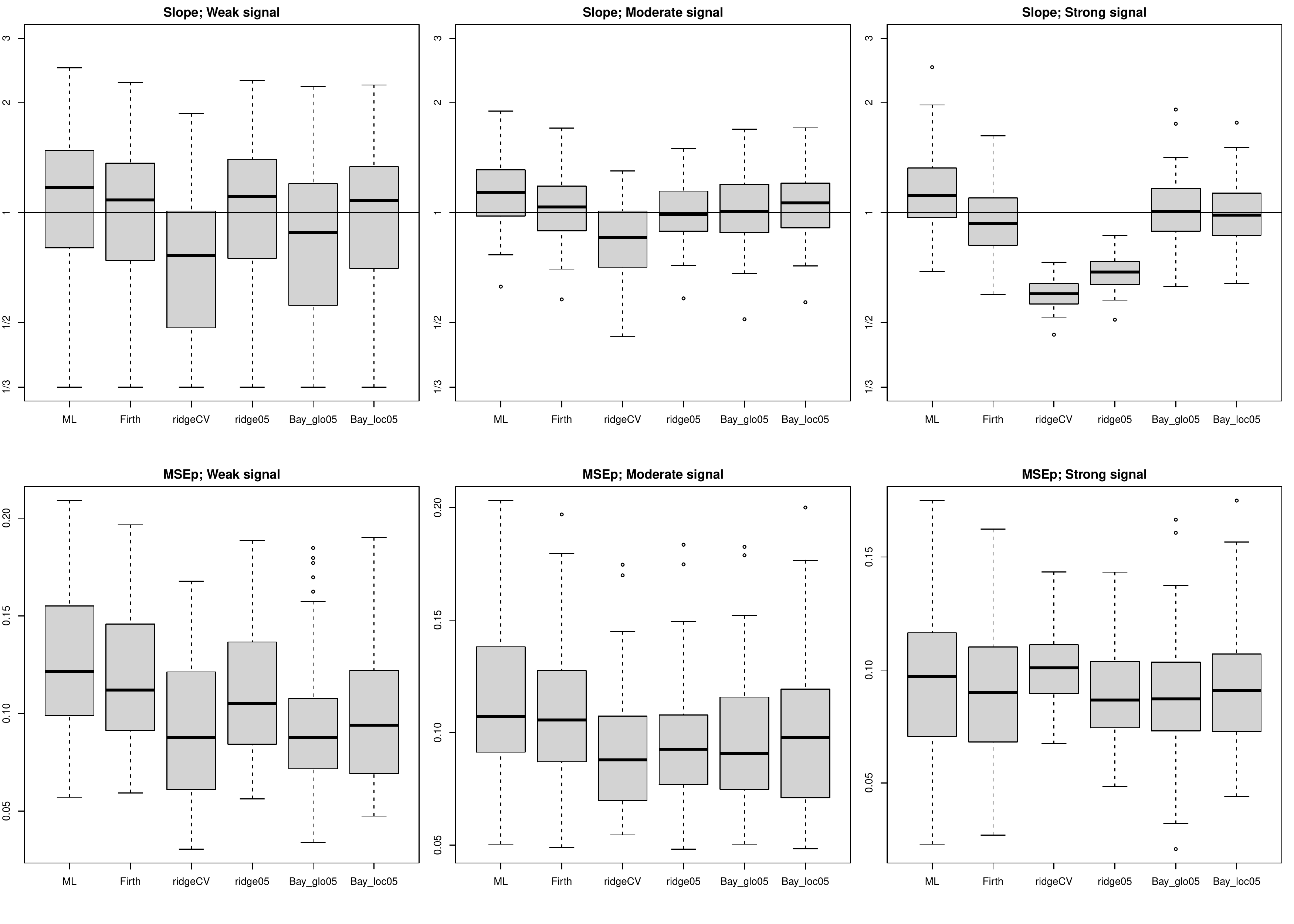}
\caption{Calibration (top; cslope: winsorized at $(1/3,3)$, log-scale) and prediction accuracy (bottom; MSEp) for logistic regression simulations. Simulation setting as described, but with sample size $n$ doubled to 100}\label{logisticn100}
\end{figure}

\begin{figure}[h]
\centering
\includegraphics[scale=0.5]{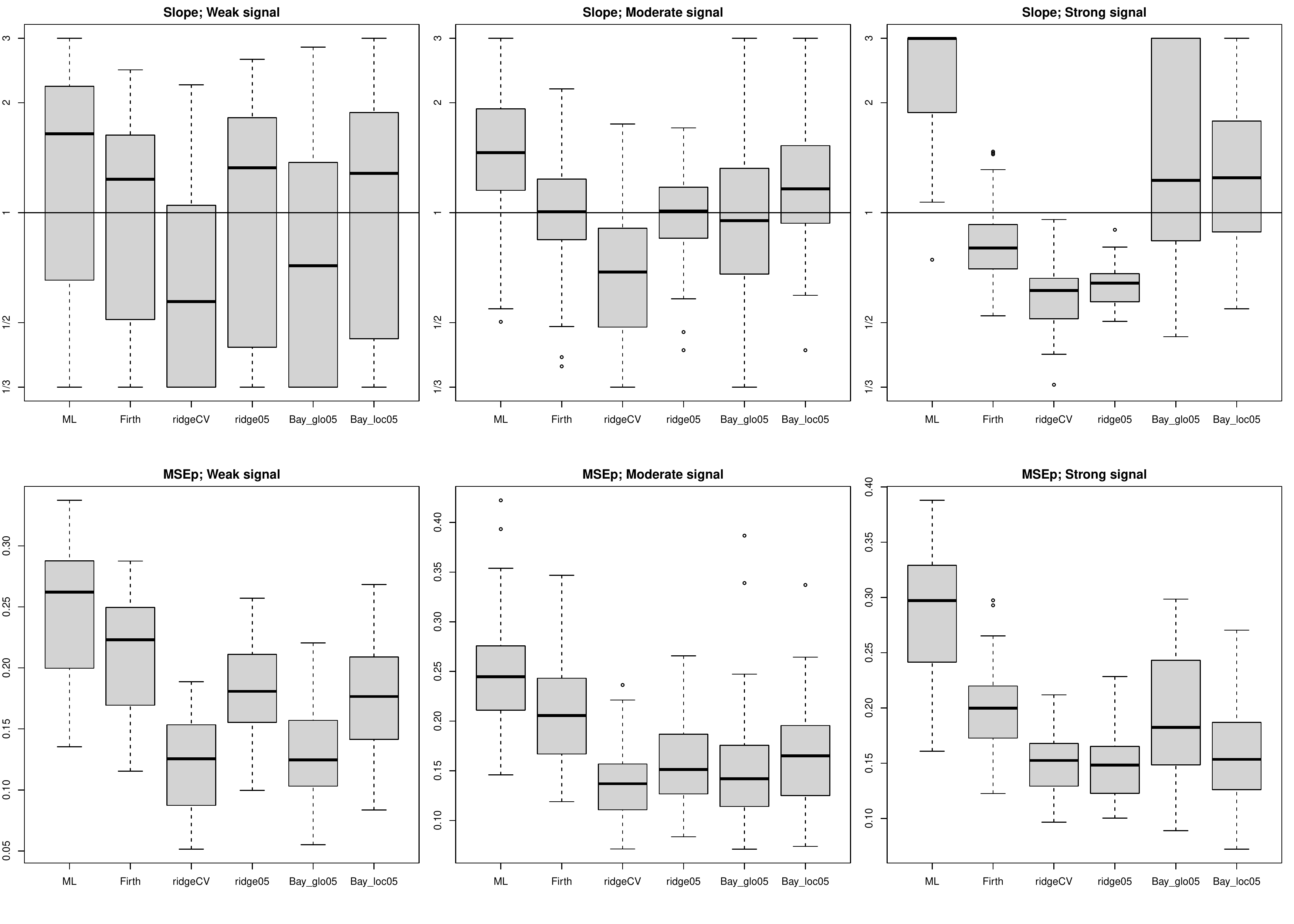}
\caption{Calibration (top; cslope: winsorized at $(1/3,3)$, log-scale) and prediction accuracy (bottom; MSEp) for logistic regression simulations. Simulation setting as described, but with
 five $\beta_j$'s added for which $\beta_j = 0$}\label{logisticaddzeros}
\end{figure}

\begin{figure}[h]
\centering
\includegraphics[scale=0.5]{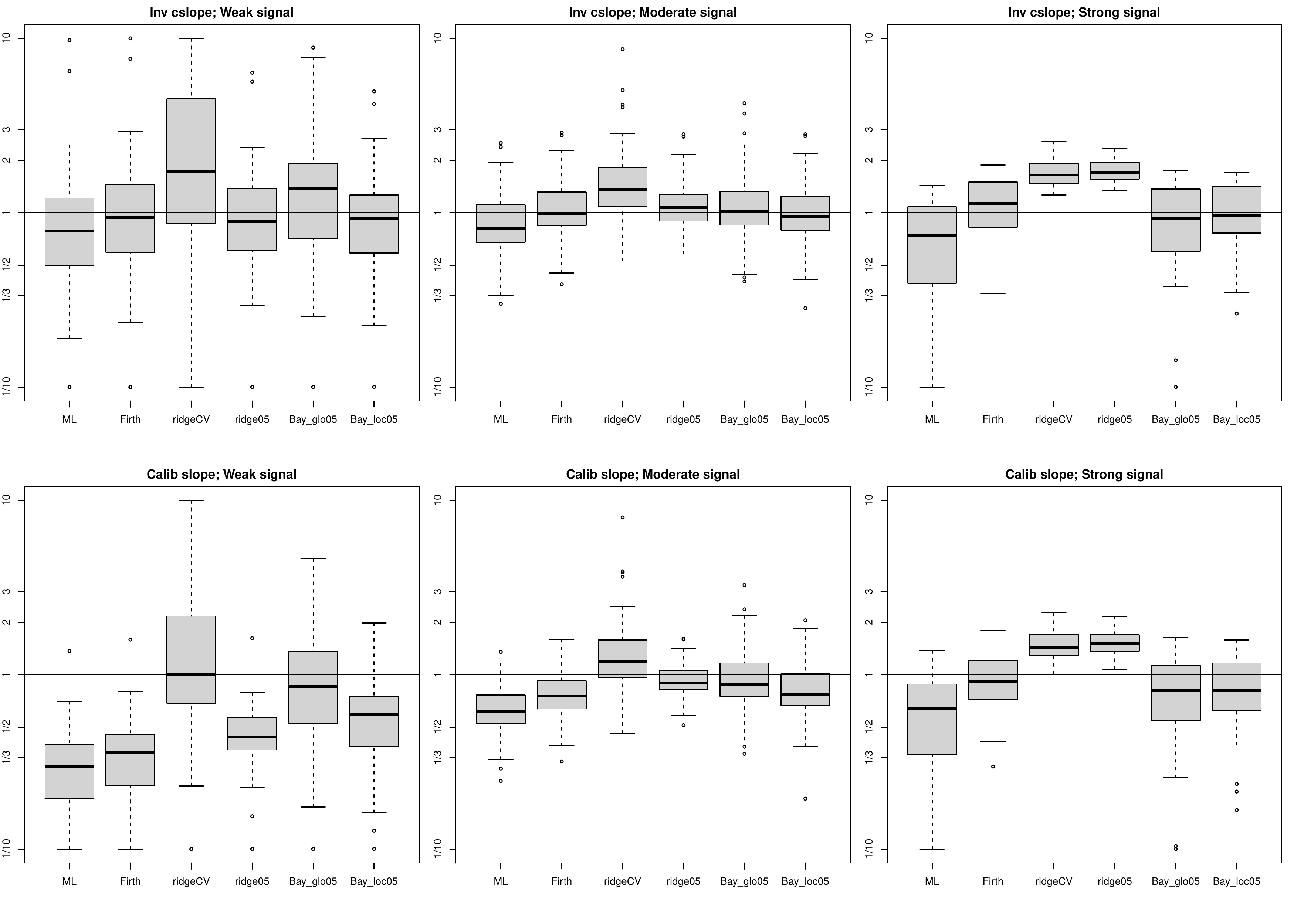}
\caption{Inverted cslope (top) and calibration slope (bottom) for logistic regression simulations; both winsorized at $(1/10,10)$ and on log-scale.
Simulation setting as in main document. As the predictions are used as dependent variables for the cslope and as independent ones for the calibration slope,
the inverted cslope is displayed}\label{cslopecalslope}
\end{figure}

\begin{figure}[h]
\centering
\includegraphics[scale=0.46]{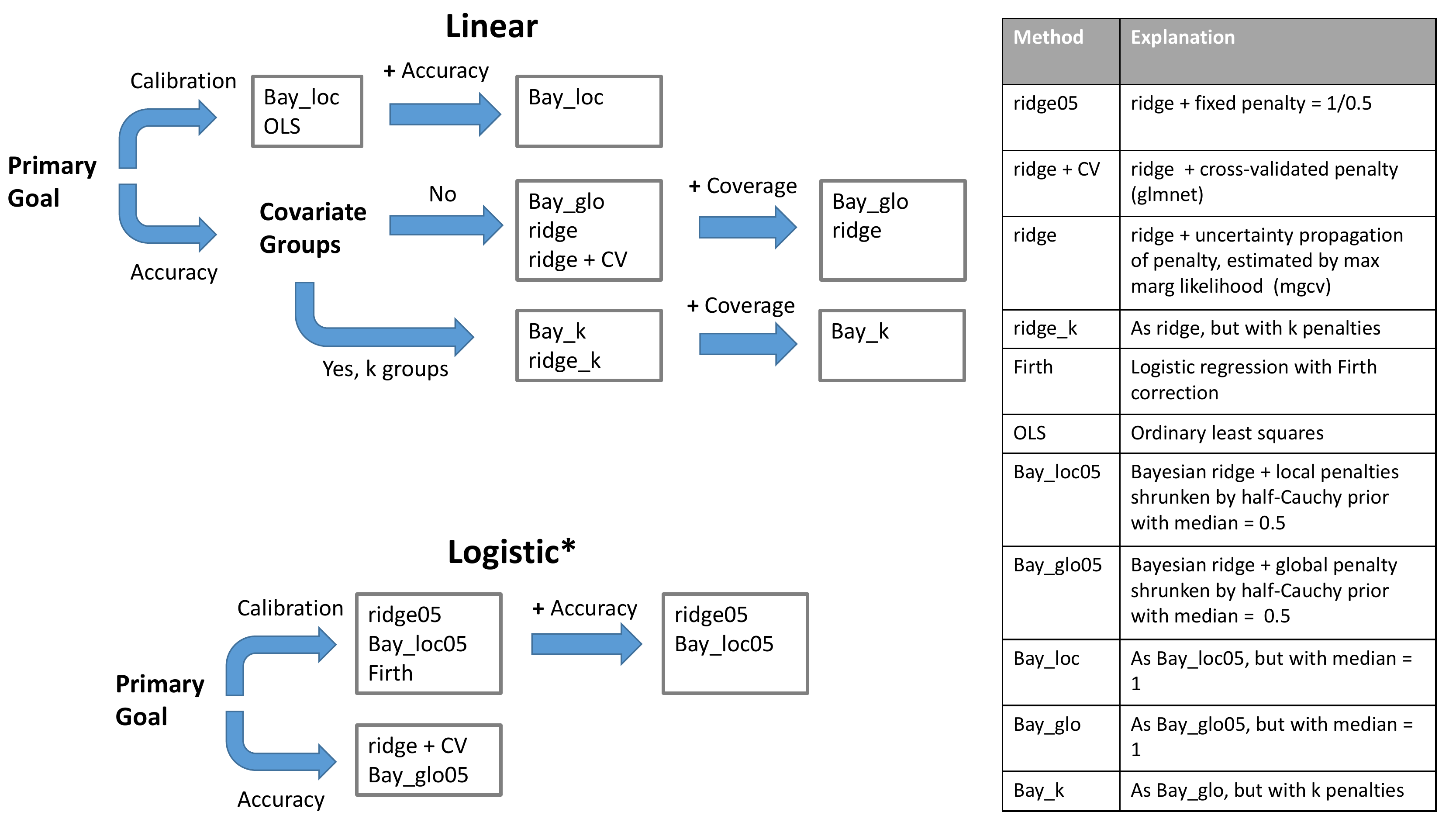}
\caption{Flow chart for recommended use of various methods, depending on primary goal (Calibration / Accuracy) and secondary goal (+ Accuracy / + Coverage).\\ $^*$Coverage and effect of covariate groups have not been studied for the logistic setting, so these are not included in this flow chart}\label{flowchart}
\end{figure}

*


\end{document}